\documentclass[10pt,twocolumn,showpacs,amsmath,amssymb,floatfix,superscriptaddress]{revtex4-2}

\usepackage{graphicx}
\usepackage{dcolumn}
\usepackage{bm}
\usepackage{color}
\usepackage{txfonts}
\usepackage{microtype}
\usepackage{hyperref}

\usepackage{easyReview}

\setreviewson

\begin{document}

\title{Enhanced signature of vacuum birefringence in a plasma wakefield}

\author{Feng Wan}
\affiliation{Ministry of Education Key Laboratory for Nonequilibrium Synthesis and Modulation of Condensed Matter, Shaanxi Province Key Laboratory of Quantum Information and Quantum Optoelectronic Devices, School of Physics, Xi'an Jiaotong University, Xi'an 710049, China}
\author{Ting Sun}
\affiliation{Ministry of Education Key Laboratory for Nonequilibrium Synthesis and Modulation of Condensed Matter, Shaanxi Province Key Laboratory of Quantum Information and Quantum Optoelectronic Devices, School of Physics, Xi'an Jiaotong University, Xi'an 710049, China}
\author{Bai-Fei Shen}
\affiliation{Department of Physics, Shanghai Normal University, Shanghai 200234, China}
\author{Chong Lv}
\affiliation{Department of Nuclear Physics, China Institute of Atomic Energy, P. O. Box 275(7), Beijing 102413, China}%
\author{Qian Zhao}
\affiliation{Ministry of Education Key Laboratory for Nonequilibrium Synthesis and Modulation of Condensed Matter, Shaanxi Province Key Laboratory of Quantum Information and Quantum Optoelectronic Devices, School of Physics, Xi'an Jiaotong University, Xi'an 710049, China}
\author{Mamutjan Ababekri}
\affiliation{Ministry of Education Key Laboratory for Nonequilibrium Synthesis and Modulation of Condensed Matter, Shaanxi Province Key Laboratory of Quantum Information and Quantum Optoelectronic Devices, School of Physics, Xi'an Jiaotong University, Xi'an 710049, China}
\author{Yong-Tao Zhao}
\affiliation{Ministry of Education Key Laboratory for Nonequilibrium Synthesis and Modulation of Condensed Matter, Shaanxi Province Key Laboratory of Quantum Information and Quantum Optoelectronic Devices, School of Physics, Xi'an Jiaotong University, Xi'an 710049, China}
\author{Karen Z. Hatsagortsyan}\email{k.hatsagortsyan@mpi-hd.mpg.de}
\affiliation{Max-Planck-Institut f\"{u}r Kernphysik, Saupfercheckweg 1,
69117 Heidelberg, Germany}
\author{Christoph H. Keitel}
\affiliation{Max-Planck-Institut f\"{u}r Kernphysik, Saupfercheckweg 1,
69117 Heidelberg, Germany}
\author{Jian-Xing Li}\email{jianxing@xjtu.edu.cn}
\affiliation{Ministry of Education Key Laboratory for Nonequilibrium Synthesis and Modulation of Condensed Matter, Shaanxi Province Key Laboratory of Quantum Information and Quantum Optoelectronic Devices, School of Physics, Xi'an Jiaotong University, Xi'an 710049, China}
\date{\today}

\begin{abstract}
Vacuum birefringence (VB) is a basic phenomenon predicted in quantum electrodynamics (QED). However, due to the smallness of the signal, conventional magnet-based and extremely intense laser-driven	detection methods are still very challenging. This is because in the first case the interaction length is large but the field is limited, and vice versa in the second case. We put forward a method to generate and detect VB in a plasma bubble wakefield, which combines both advantages, providing large fields along large interaction lengths. A polarized $\gamma$-photon beam is considered to probe the wakefield along a propagation distance of millimeters to centimeters in the plasma bubble. We find via plasma particle-in-cell simulations  that the VB signal in terms of Stokes parameters can reach about $ 10^{-5}$ ($10^{-3}$-$10^{-2}$) for tens of MeV (GeV) probe photons  with moderately intense lasers ($10^{20}$-$10^{21}~\mathrm{W/cm^2}$). The main source of noise from plasma electrons is mitigated, in particular, by a choice of $\gamma$-photon polarization and by proper modulation of the plasma density. The proposed method represents an attractive alternative for the experimental observation of VB via laser-plasma interaction.
\end{abstract}
\maketitle

In quantum electrodynamics (QED)  the vacuum is fluctuating with the virtual creation and annihilation of electron-positron pairs due to the Heisenberg uncertainty principle which, in particular, leads to the vacuum polarization  \cite{Halpern_1933,Euler_1935,Schwinger_1949_Quantum}.
The latter in external electromagnetic fields transforms the vacuum into a refractive medium \cite{Schwinger_1951_Gauge,Klein_1964_Birefringence,Brezin_1971_Polarization,XiePRD}, and the anisotropy induced by the strong field renders the vacuum birefringent \cite{dittrich2000probing}.
As one of the basic  predictions of nonlinear QED, the vacuum birefringence (VB) effect is a covet for experimental observation, and in recent years it has been extensively studied in various strong-field configurations; see  e.g.~\cite{Marklund2006,Piazza2012,Karbstein_2013_Photon,Gonoskov2021}.

The experimental attempts, though yet not successful, to measure the VB effect have been mostly connected with the effect in a static magnetic field of a stationary magnet in the laboratory (PVLAS project \cite{Ejlli2020}, and its modifications: BFRT \cite{Cameron1993}, and BMV \cite{Cadene2014}), which provides a field strength of the order of Tesla within a rather large scale (about centimeters to meters). The VB signal of an optical probe wave as a rotation angle of the wave polarization amounts to $\delta \phi \sim 10^{-11}$-$10^{-10}$, which is still beyond the current accuracy of measurements (about $10^{-9}$ for optical photons) \cite{Ejlli2020}. Here, $\delta \phi \propto B_\perp^2 L / \lambda_\gamma$, where $L$ is the total distance of the probe photon traversing the transverse magnetic field $B_\perp$, and $\lambda_\gamma$ the wavelength of the probe photon.
As the VB signal is proportional to the energy of the probe photons, there are proposals to use high-energy $\gamma$-photons \cite{Cantatore_1991_Proposed}, e.g., tens of GeV $\gamma$-photons traversing through a several-meter-long large-scale superconducting magnet, where $\delta \phi$ may reach $10^{-4}$ \cite{Wistisen_2013_Vacuum}. However, this would demand a $\gamma$-photon beam with a challenging high collimation.  While the VB effect is magnified via multiple reflections in a magnetized cavity, the latter introduces considerable systematic errors.

With advancement of strong laser technique in recent years \cite{Gales_2018_extreme,Danson_2019_Petawatt,Yoon_2019_Achieving,Yoon_2021}, there are justified hopes to detect the VB effect with laser fields, being the largest fields in a laboratory. In fact, the lasers provide $\sim 10^5$ times larger fields than via laboratory static magnets. The main disadvantage of the setup with ultrastrong laser fields is that the high field interaction region is limited to the focal size of the laser beam of several micrometers.  VB schemes with the use of lasers have been widely investigated theoretically, with probe photon energies spanning from eV (optical photons) \cite{DiPiazza_2006,Heinzl_2006_observation,Piazza2007Enhancement,Klar2020}, tens of keV (X-rays) \cite{Schlenvoigt_2016_Detecting,Karbstein_2016lby,King_2016_Vacuum,Shen_2018,Seino2020} to GeV ($\gamma$-rays) \cite{Bragin_2017,Nakamiya_2017,Karbstein2021}.
Since the pulse duration of multi-PW lasers is limited to tens of femtoseconds, for keV X-ray photons, the signal may reach $\delta\phi \simeq 10^{-5}$. Owing to the low flip rate of polarization and instability of PW laser facilities ($\sim 20\%$ \cite{Yoon_2021}), the signal measurements usually require $10^2$-$10^3$ shots to accumulate sufficient statistics \cite{Schlenvoigt_2016_Detecting}.
For GeV probe $\gamma$-photons the polarization flip rate is much higher ($10^{-3}$-$10^{-1}$), but the final signal is accompanied by electron-positron pair production (vacuum dichroism), and  thousands of shots are required to reach a high confidence level \cite{Bragin_2017, Nakamiya_2017}. Thus, the VB effect is still not verified directly in experiments and, an efficient, stable, and compact detection method is still in great demand.

\begin{figure*}[!t]
	\centering
	\includegraphics[width=0.7\linewidth]{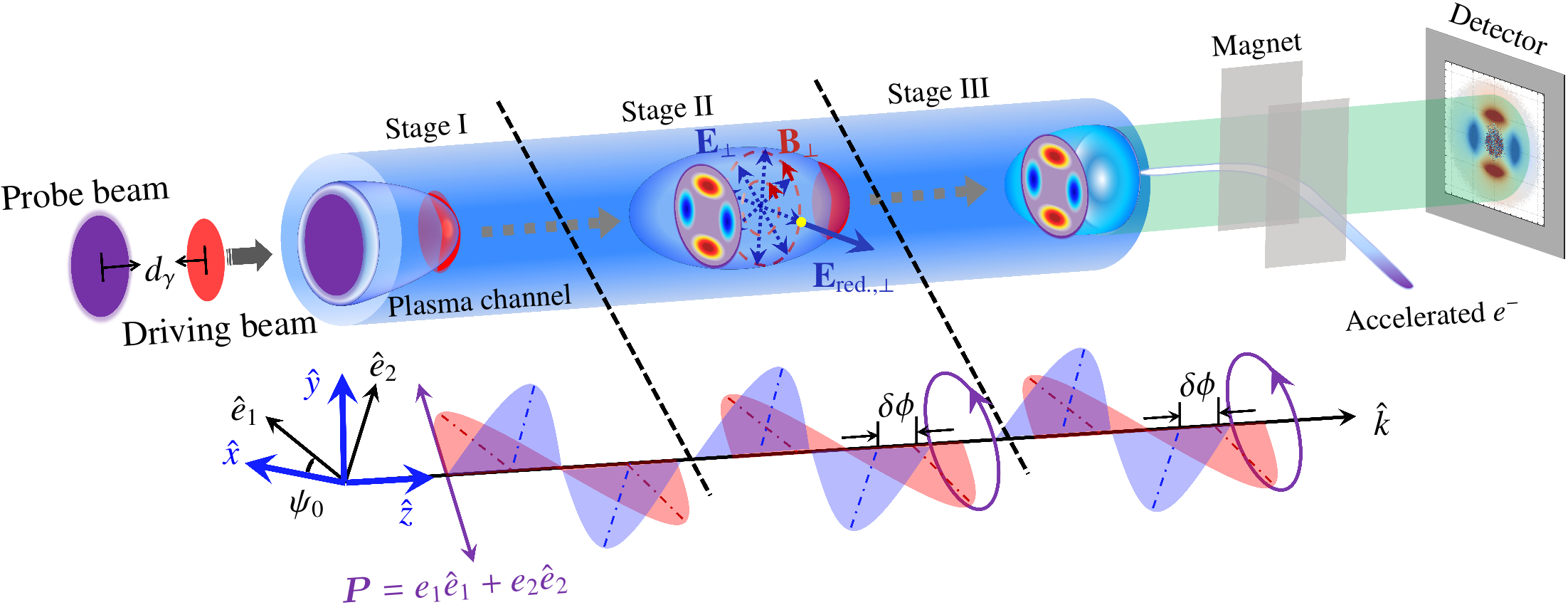}
	\caption{Interaction scenario.
Upper column:  the formation process of the VB signals in a plasma bubble.
Lower column: the schematic evolution of the polarization vector of a probe photon in the bubble wakefield.
	} \label{fig1}
\end{figure*}

\begin{figure}[!t]
	\includegraphics[width=\linewidth]{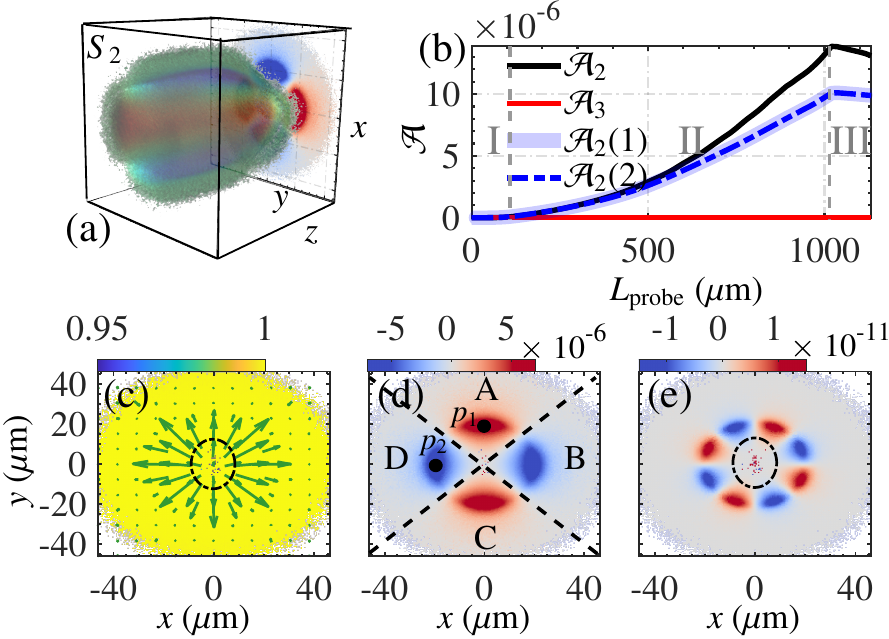}
	\caption{Signatures of the VB effect with a LP $\gamma$-photon probe: (a) 3D distribution of $\overline{S_2^f}$ of injected probe $\gamma$ photons; (b) evolution of the asymmetry parameters $\mathcal{A}_i$, where $i = 2, 3$ refer to $S_{2,3}^f$, respectively; black and red lines are calculated with photons from cones centered at $p_1$ and $p_2$ with both radii of $r = 1~\mathrm{\mu m}$. Light-blue solid and blue dash-dotted lines are calculated by including and excluding the NCS photons (in ``A''-``D'') with the transverse position $r \geq 10~\mathrm{\mu m}$, respectively; (c), (d), and (e): the Stokes parameters $S_1^f$, $S_2^f$ and $S_3^f$ of $\gamma$-photons with $\omega_{\gamma} \in 20\pm5$~MeV and azimuthal angle $\Delta \theta \leq 10$~mrad [including radiated photons (within the black dash-dotted circles) and injected probe photons] in the transverse $x$-$y$ plane (averaged along $z$ direction), respectively. Green arrows in (c) indicate the orientation of the transverse reduced field $\mathbf{E}_\mathrm{red \perp}$, while $p_1$ at ($-19.5~{\rm \mu m}, 0$) and $p_2$ at ($0, 19.5~{\rm \mu m}$) denote two cylinders along $\hat{z}$ with both radii of $r$. Other parameters are given in the text.} \label{fig2}
\end{figure}

In this Letter, we put forward such an efficient and compact scheme for a VB measurement employing currently available experimental techniques. In this scheme, a laser beam of moderate-intensity ($10^{20}$-$10^{21}~\mathrm{W/cm^2}$) generates a bubble wakefield in plasma. The VB induced by the ultrastrong field of the plasma bubble is probed by  $\gamma$-photons which propagate in plasma synchronized with the bubble over a long distance; see the interaction scenario in Fig.~\ref{fig1}. The whole interaction process can be divided into three stages.
In stage I, a driving beam excites a  wakefield in plasma in the bubble regime. In stage II, linearly polarized (LP) $\gamma$-photons are injected into the plasma bubble experiencing the stable and ultrastrong transverse electromagnetic field of the bubble. Due to VB in the ultrastrong field region, a small circular polarization (CP) of $\gamma$-photons arises determined by the phase retardation $\delta \phi$. The characteristic signature of VB may form an angular pattern of a four- (eight-)leaf clover (see Figs.~\ref{fig1} and \ref{fig2}).
In stage III, the plasma bubble begins to extinguish, and the probe photons are extracted and detected. We have carried out
three-dimensional (3D) QED particle-in-cell (PIC) simulations of the plasma and tracked $\delta \phi$ for $\gamma$-photons, taking into account the explicit plasma field along the $\gamma$-ray. The effective transverse field in the bubble is shown to reach the order of  $10^3$-$10^4$ T, within the interaction region of $1$ cm, allowing for $\delta \phi$  up to about $10^{-5}$ ($10^{-3}$-$10^{-2}$), with tens of MeV (GeV) probe photons and moderately intense laser fields. The main limitation of the scheme is the noise because of plasma electrons.
The Compton scattering (CS) of probe photons and the plasma emission in a strong field are included in our simulations and taken into account in the evaluation of the VB signal. The first effect appears to be not significant because of the judiciously chosen linear polarization for the probe, when the small CP of the $\gamma$-photons during interaction can arise only due to VB but not due to CS \cite{Landau_4}. The effect of the plasma radiation, which is limited in spectral range,  is also shown not to distort significantly the VB signal.

Let us introduce our simulation method. Relativistic units with $c=\hbar=1$ are used throughout.
The VB effect of probe photons in a bubble wakefield is described by the polarization flip of the Stokes parameters
$
	\bigg(
	\begin{array}{c}
		S_1^f \\
        	S_2^f \\
		\end{array}
		\bigg) = \bigg(
    \begin{array}{cc}
		\cos\delta\phi & -\sin\delta\phi \\
		\sin\delta\phi & \cos\delta\phi
	\end{array} \bigg) \bigg(
	\begin{array}{c}
		S_1^{i} \\
		S_2^{i}
    \end{array} \bigg),
    \label{stokes}
$
where $S^i = \left(S_0^i,S_1^i,S_2^i,S_3^i\right)$ and $S^f = \left(S_0^f,S_1^f,S_2^f,S_3^f\right)$ are the initial and final Stokes parameters of the probe photons, respectively, $S_1$ and $S_3$ indicate the linear polarization, $S_2$ the CP, and $S_0$ the intensity \cite{McMaster1961}. The VB effect is determined by the phase retardation between two eigenstates of polarization  $\delta \phi = \int_0^{L} {\rm d}l \frac{2\pi}{\lambda_\gamma} \Delta n(\omega_\gamma)$,
with $\Delta n = n_\parallel - n_\perp$,  with $\parallel, \perp$ denote the polarization modes parallel to the two polarization eigenstats $\hat{e}_1$ and $\hat{e}_2$, respectively, $\hat{e}_1 \parallel {\bf E}_{\rm red \perp}$, $\hat{e}_2 = \hat{k} \times \hat{e}_1$, $\mathbf{E}_{\rm red \perp} \equiv (\mathbf{E} + \hat{k}\times\mathbf{B})_\perp$ is the transverse reduced field, and $\hat{k} $ the unit vector along the propagation direction of the probe photon. 
The VB refractive index $n(\omega_\gamma)$ in an as here sufficiently slowly varying bubble wakefield can be treated
as in a constant field. To suppress pair production and the consequent vacuum dichroism, we limit the quantum nonlinearity parameter to $\chi_\gamma = |e|\sqrt{-(F_{\mu \nu}k_\gamma^\nu)^2}/m_e^3 \ll 1$ ($S_0$), with the electron charge $-e$, mass $m_e$, frequency $\omega_\gamma$, four wave vector $k_\gamma^\mu$ of the probe photon,  field tensor $F_{\mu \nu}$, and use a simple expression for $n(\omega_\gamma)$ (see Sec. I of \cite{supplemental} and Refs. \cite{Schubert_2000_vacuum, Shore_2007_Superluminality, wolfram}) which is identical to that of the low-frequency limit \cite{Baier_1967_vacuum,Dinu2014}:
\begin{eqnarray}
	{\rm Re}[n(\omega_\gamma)]
	= 1+\frac{\alpha}{90\pi}\overline{\chi}^2_\gamma
	\left\{
		\begin{array}{c}
			4_\parallel \\
        		7_\perp \\
		\end{array}
	\right\}, \label{refractive1}
\end{eqnarray}
where $\overline{\chi}_\gamma \equiv \chi_\gamma / \omega_\gamma$. 
The VB effect via the phase delay $\delta\phi$ has been implemented into the 3D PIC code EPOCH \cite{Arber2015,Xue2020}; see details in Sec. III of \cite{supplemental}.

The signatures of the VB effect in the wakefield and its development during the interaction are illustrated in Fig.~\ref{fig2}. Here we use as the driving beam a 10-cycle LP laser pulse with the invariant field parameter $a_0=|e|E_0/m_e \omega_0=40$, the wavelength $\lambda_0=0.8~\mu$m (with the corresponding laser peak intensity $I_0\simeq 1.37 \times 10^{18} a_0^2 \left(\frac{1~\mu{\rm m}}{\lambda_0}\right)^2~{\rm W/cm^2}\approx 3.44 \times 10^{21}~{\rm W/cm^2}$), frequency $\omega_0 = 2\pi/\lambda_0$, and the focal radius $w_0 = 12~\mu\mathrm{m}$ (the cases of $a_0= 5, 10$, and 20 are discussed in Sec. V of \cite{supplemental}). The moderate laser intensity is beneficial for noise suppression. The background plasma is pure hydrogen gas plasma, which can be generated by the field ionization using the driving laser beam \cite{Wang_2021_Free}, or by capillary discharge \cite{Gonsalves2019}. The plasma  density $n_e$ linearly increases from zero (at $z=0$) to $0.002n_c$ (at $z = 100~\mu\mathrm{m}$) and then to $0.007n_c$ (at $z = 1.0~$mm), where $n_c = m_e\omega_0^2/4\pi e^2$ is the critical plasma density (other cases with different plasma densities are discussed in Sec. V of \cite{supplemental}).
The probe beam is composed of $10^7~\gamma$-photons with an average energy of $\varepsilon_\gamma$ = 20~MeV, energy spread $\Delta\varepsilon_\gamma$ = 1~MeV, and angular spread $\Delta\theta = 10$~mrad [such $\gamma$-beam can be delivered by ELI-NP within several years \cite{Ur_2015}, single-shot all-optical or beam-foil nonlinear Compton scattering (NCS) \cite{Li_2020_Polarized, Xue2020, Sampath_2021}]. The spatial distribution of  $\gamma$-photons is a Gaussian with a longitudinal radius $r_l = 7~\mu$m, and transverse radius $r_t = 15~\mu$m.
The probe photon beam propagates along the $\hat{z}$ direction and is LP with the Stokes parameters $S^i=(1, 1, 0, 0)$ (the case of circularly polarized probe is discussed in Sec. IV of \cite{supplemental}). The simulation box of the moving window is set as  $\Delta x \times \Delta y \times \Delta z$ = $90~\mu{\rm m}\times90~\mu{\rm m}\times100~\mu{\rm m}$, with the spatial grid size $180\times 180 \times 1600$. The macro-particles per cell are 5 and 1 for electron and proton, respectively, and the total number of macro-particles for probe $\gamma$-photons is set as $1.5\times 10^6$.
To ensure that probe photons can experience strong fields and stay in the bubble as long as possible, the probe beam is synchronized with the driving beam with a relative delay of $d_\gamma = 27~\mu$m (cf. with the radius of the bubble  $\rho_b \simeq (2/k_p)\sqrt{a_0} \simeq 20~\mu{\mathrm m}$, where $k_p =  \sqrt{4\pi n_e e^2 / m_e}$ and $n_e \simeq 0.01n_c$ \cite{supplemental, Esarey2009}).

For the initially LP probe, a final CP emerges due to VB, with $S_2^f\simeq 7 \times10^{-6}$, at the interaction length $1$~mm. The spatial distribution of $S_2^f$ forms a unique  structure of a four-leaf clover (eight leaves for $S_3^f$) [Figs.~\ref{fig2}(a), (d), (e)], while the change in $S_1$ is negligible [Fig.~\ref{fig2}(c)]. For detection convenience, here, we define an asymmetry  parameter $\mathcal{A} \equiv \overline{S^f_2}(A)+\overline{S^f_2}(C)- \overline{S^f_2}(B)-\overline{S^f_2}(D)= |\overline{S^f_2}(A)|+|\overline{S^f_2}(B)|+|\overline{S^f_2}(C)|+|\overline{S^f_2}(D)|$, where $\overline{S^f_2}$ is the averaged Stokes parameter in regions ``A''-``D'' with each one containing a leaf of the clover in Fig.~\ref{fig2}(d).
Compared with PVLAS, the plasma wakefield can maintain a stronger field $B \gtrsim 10^3$ T, and the corresponding VB signal $B^2L \simeq 10^{3}$-$10^{4}$ ${\rm T^2  m}$,  will be one to two orders larger than in the planned PVLAS-FE  \cite{Ejlli2020}.

\begin{figure}
	\includegraphics[width=\linewidth]{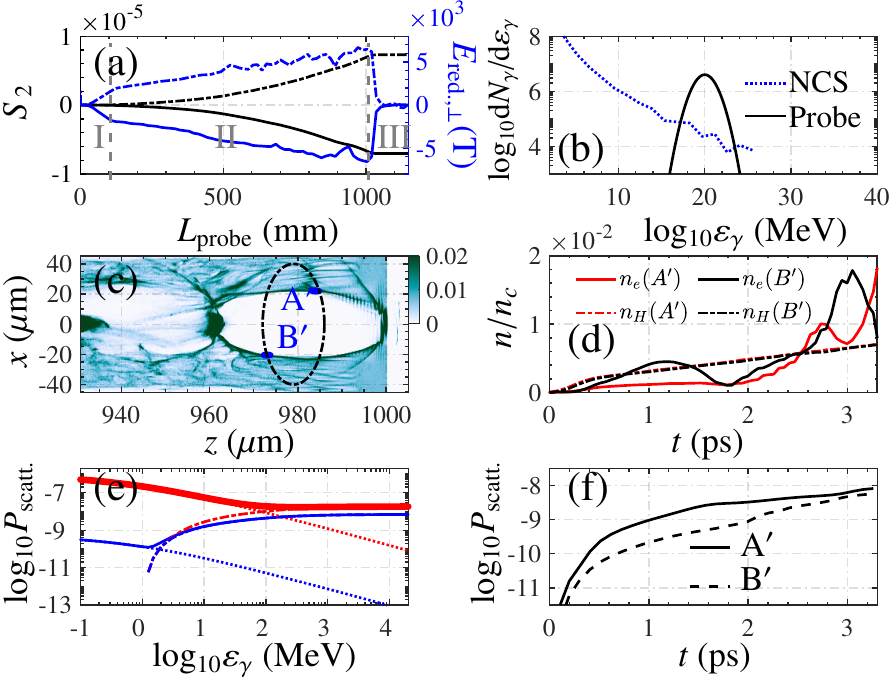}
	\caption{(a) Stokes parameter $S_2$ (black) and the corresponding experienced reduced field $\mathbf{E}_{\rm red \perp}$ (blue) vs the interaction path $L$~(mm) of two sampled particles with solid and dash-dotted lines denoting particles from $p_1$ and $p_2$ in Fig.~\ref{fig2}(d), respectively. (b) Energy spectra of the plasma emission (NCS, blue-dotted) and injected probe photons (black-solid). (c) Number density of electrons of the bubble, where the gray dash-dotted circle indicates the probe beam position. (d) The experienced plasma densities of sampled photons from ``A$'$'' and ``B$'$'' in (c), with solid and dash-dotted lines denoting electrons ($n_e$) and ions ($n_H$), respectively. (e) Scattering probability log$_{10}(P_{\rm scatt.})$ vs photon energy log$_{10}(\varepsilon_{\gamma})$~(MeV) of $\gamma$-photons traversing a hydrogen plasma of 2~mm, where blue and red lines are taken from the center ($n_e = 6\times10^{-6}n_c$) and the sheath ($n_e = 0.012n_c$) of the bubble, respectively. Dot, dash-dotted and thick solid lines indicate CS, total pair production (BH and trident pair productions) and the total scattering (CS + pair production), respectively. (f) Total scattering probabilities log$_{10}(P_{\mathrm{scatt.}})$ vs time $t$ for probe photons (20 MeV) located at ``A$'$'' and ``B$'$'', respectively, where $P_{\mathrm{scatt.}}(t) \equiv 1- \prod_{i = 1}^n \exp\left\{-\left[n_e(t_i) (\sigma_\mathrm{CS} + \sigma_{\pm, \mathrm{tri.}}) + n_H(t_i) \sigma_{\pm, \mathrm{BH}} \right] c\Delta t \right\}$, $t_i = i\Delta t$, $t=n\Delta t$ and $\Delta t$ is the time step size.} \label{fig3}
\end{figure}

Since the driving laser is LP, the low-energy X-ray from betatron radiation and $\gamma$-photons from NCS are LP, i.e., they can affect $S_1^f$ and $S_3^f$, but not CP; see signals within the black dash-dotted circles of Figs.~\ref{fig2}(c) and (e). 
Theoretically, $S_2$, i.e., the CP will not be affected, because CS of LP photons cannot produce circularly polarized photons \cite{Landau_4} as is confirmed in  Fig.~\ref{fig2}(d) (within the black dash-dotted circle) and more details in Sec. IV of \cite{supplemental}.
As the VB signal is proportional to the photon energy $|S_{2,\mathrm{LP}}^f|~\sim  \delta\phi \propto \omega_{\gamma}$, higher-energy probe photons can produce stronger signals, for instance, at $\omega_{\gamma}\sim 1$ GeV, $\delta \phi \sim 10^{-3}$, with noise still insignificant [Figs.~\ref{fig3} (e), (f)]. Thus, the CP signal of VB in the plasma wakefield setup is of the order of $|S_{2}^f| \simeq 3.5\times 10^{-6}$ and can be increased using a longer interaction length in the plasma. For the detection of such a weak CP signal of a $\gamma$-photon beam we could follow the principles of sensitive circular polarimetry methods in \cite{Schopper_1958,Fukuda_2003,Tashenov_2011,Ilie_2019}; and for the detection of the linear polarization, the polarization-dependent Bethe-Heitler (BH) pair production method can be employed \cite{Bragin_2017}.
The number of detected $\gamma$-photons ($N_\gamma$)  should be large enough to suppress the statistical error $1/\sqrt{N_\gamma}$ below the required accuracy $|S_{2}^f|\sim  3\times 10^{-6}$, i.e. $N_\gamma\sim 10^{11}$. With  $10^7$ as number of $\gamma$-photons in a beam, one will need $10^3 \sim 10^4$ shots for statistical accuracy. These requirements can be fulfilled by all-optical NCS \cite{Li_2020_Polarized, Xue2020} or beam-foil radiation \cite{Sampath_2021}.
We underline that the considered setup is realized with tabletop 100s TW or PW laser systems, which provide much more stable fields in the plasma wakefield, compared with the fields of 10-100 PW laser systems.

The development of the VB signal is elucidated in Fig~\ref{fig3}(a).  In the plasma wakefield, the magnitude of the transverse field $\mathbf{E}_{{\rm red} \perp}$ is proportional to $k_p \propto \sqrt{n_{\rm e0}}$ \cite{Kostyukov2004,Lu2006}.
In stage I, the excited wakefield in the low-density plasma region is rather weak, and the VB effect is negligible.
In stage II, due the applied density gradient, $|\mathbf{E}_{\rm red \perp}|$ linearly rises
up to $|\mathbf{E}_{\rm red \perp}| \simeq 6000$ T (at $z \simeq 1~{\rm mm}$), and the phase retardation of probe photons increases as $\delta\phi \propto |\mathbf{E_{\rm red \perp}}|^2L \propto L^3$, yielding the VB signal $S^{f}_2 \simeq   \frac{\alpha}{45}  \frac{|\mathbf{E}_{\rm red \perp}(1~\mathrm{mm})|^2}{E_{cr}^2} \frac{L}{\lambda_\gamma}$ amounting to $ 5\times  10^{-6}$ at $L=1$~mm [cf. Fig.~\ref{fig3}(a)], where $E_{cr}=m^2/e$ is the QED critical field.
In stage II, $\mathbf{E}_{\rm red \perp}$ is radially aligned [see Fig.~\ref{fig2}(c)], i.e., $\mathbf{E}_{\rm red \perp} \parallel (\cos\theta \hat{x} + \sin\theta \hat{y})$, with the azimuthal angle $\theta  = \arctan(y/x)$.
Defining eigenstates $\hat{e}_1 = (\cos\theta\hat{x} + \sin\theta\hat{y})$ and $\hat{e}_2 = \hat{k} \times \hat{e}_1 = (-\sin\theta\hat{x} + \cos\theta\hat{y})$, the initial Stokes parameters in this frame are $S^{{\rm LP},i} = (1, \cos2\theta, 0, \sin2\theta)$ and $S^{{\rm CP}, i} = (1, 0, 1, 0)$; see in Sec. II of \cite{supplemental}.
After traversing the polarized vacuum, the final Stokes parameters are given by $S^{{\rm LP},f} = \bigg[1, \frac{\cos4\theta + 1}{2}(\cos\delta\phi - 1) + 1, \cos2\theta\sin\delta\phi, \frac{\sin4\theta}{2}(\cos\delta\phi-1)\bigg]$ (see Sec. II in \cite{supplemental}). Therefore, $S^{{\rm LP},f}_2  =\cos2\theta\sin\delta\phi \propto \cos2\theta \cdot \rho^2\exp(-2 \rho^2/\rho_b^2)$, where we use $\delta\phi \propto |\mathbf{E}_\perp(\mathbf{B}_\perp)|^2 \propto \exp(-2\rho^2/\rho_b^2)\rho^2$, phenomenologically describing the local transverse bubble field as $\propto \exp(-\rho^2/\rho_b^2)\boldsymbol{\rho}$, with $\boldsymbol{\rho} = x\hat{x} + y\hat{y}$, and $\rho = |\boldsymbol{\rho}|$. Thus, the VB signature forms a unique structure  with four isolated poles due to the term $\cos2\theta$, with a maximum at $\rho=\rho_b$, which is consistent with Figs.~\ref{fig2}(a) and (d).
The interaction length with the bubble structure is limited by the dissipation of the laser energy, $L_{dp} \lesssim \left(\frac{n_e}{n_c}\right)^{-3/2} \lambda_0 \frac{\sqrt{2}}{\pi}a_0 \simeq 1.8$~cm for $n_e = 0.01n_c$ \cite{Esarey2009} ($S_2^f \sim 6 \times 10^{-5}$, and only 25 shots are required to suppress the statistical error),
but the wakefield scales as ($|\mathbf{E}| \propto \sqrt{n_e}$),
such that for the VB signal $\delta \phi \propto |\mathbf{E}|^2 L_{pd} \propto \frac{a_0}{\sqrt{n_e}}$, an intense driver with low-density plasma is beneficial.

\begin{figure}
	\includegraphics[width=\linewidth]{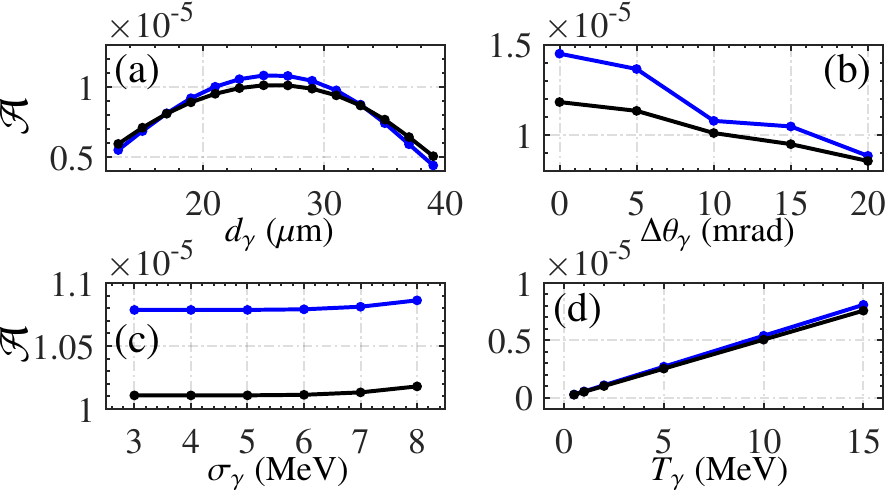}
	\caption{Impact of the probe $\gamma$-photon beam parameters on the VB signal via the parameter $\mathcal{A}$ vs: (a) spatial delay $d_\gamma$,  (b) angular spread $\Delta\theta$, (c) energy spread $\sigma$, and (d) temperature $T$ of the probe beam. In (a) and (b), the beam energy is assumed to follow a Gaussian distribution with the energy spread of 1 MeV. In (a), (c) and (d), the angular spread is 10~mrad. The probe beam energy in (c) is a Gaussian distribution $f(\omega; \omega_0, \sigma) \propto \frac{1}{2\pi\sigma}\exp\left(-\frac{(\omega - \omega_0)^2}{2\sigma^2}\right)$ with the central energy $\omega_0$ and RMS $\sigma$, respectively, and in (d) a Maxwellian distribution $f(\omega) \propto \exp\left(-\omega/T\right)$ with temperature $T$. Black (Blue) lines indicate the signal collected from photons with transverse position $r \geq 10~(15.7)~\mathrm{\mu m}$. All other parameters are the same as those in Fig.~\ref{fig2}.} \label{fig4}
\end{figure}

In the considered setup noise from plasma electrons may affect the VB signal. Such noise stems from:
1) self-generated photons of the plasma electrons, and 2) CS of the probe photons off the plasma electrons. Both the self-generated photons and the scattered ones  mix with the probe and thus affect the final VB signal. During propagation of the wakefield, electrons are self-injected into the wakefield, accelerated to high energies,  
and can emit (low-energy X-rays via betatron radiaiton \cite{Corde2013} do not affect the VB signal) high-energy photons via NCS in the strong external field. These newborn photons via NCS are LP \cite{King2020,Tang2020,Li_2020_Polarized} and they do not change the number of circularly polarized photons generated via VB, however, they would reduce the average CP of the photon beam impinging on the detector. Nevertheless, the radiation spectrum of the NCS peaks around $ \omega_{\rm NCS} \sim 0.44 \gamma_e m_e \chi_e \ll \omega_{\rm probe}$ as $|\mathbf{E}_{\rm red \perp}| \simeq 6000$ T, $\gamma_e \lesssim 2000$ and $\chi_e \lesssim 0.001$ \cite{Ritus_1985_Quantum,Baier_1998_Electromagnetic,Bell_2008_Possibility,Ridgers_2012_Dense}; see the energy spectra of radiated photons in Fig.~\ref{fig3}(b), and they are naturally separated from the probe beam in angular distribution [Figs.~\ref{fig2}(c)-(e)].
The applied setup with a linearly increasing smooth gradient also suppress the self-injection of electrons and subsequent NCS.

As for the second problem, the high-energy probe photons ($\omega_\gamma \gg 2m_e$) propagating in the hydrogen plasma wakefield, can scatter off background electrons  (CS) and produce pairs in the electron (trident process) and proton field (BH process \cite{nist-xcome, supplemental, Tsai_1974_Pair,Hubbell_2006_Review,Salvat_2009_Overview,Allison_2016}), and in the strong wakefield via the nonlinear Breit-Wheeler process. The pair production process could deplete the probe beam, however, the probability at the given conditions with $\chi_\gamma\sim 0.001$ is $P_\pm \lesssim 10^{-8}$, and the probe depletion due to pair production is negligible; see Sec. IV of \cite{supplemental}.

The CS of LP probe photons cannot create CP for the scattered photon \cite{Landau_4}, however, it could also deplete the probe beam. The probability of CS for a probe photon is maximal when it traverses the bubble sheath [where the electron number density reaches the maximum  $n_e^{(max)} \simeq 10^{-2}n_c$] and can be estimated as $P_{\rm CS} \simeq n_e \sigma_{\rm cs}  L \simeq 3 \times 10^{-8}$, for $L = 2$ mm; see Figs.~\ref{fig3}(d) and (e). 
This estimation is consistent with the real-time calculated scattering probabilities of the two sampled photons at ``A$'$'' and ``B$'$''; see Fig.~\ref{fig3}(f).

For the experimental feasibility, we also study the impact of the spatial delay $d_\gamma$, angular divergence $\Delta \theta$, energy spread, and transverse radius of the probe beam on the VB signatures.
Since the maximum distance a $\gamma$-photon can propagate in the wakefield is given by $L_\mathrm{max} \simeq \frac{c}{c-v_\mathrm{w}}d_\gamma$, where $v_\mathrm{w}$ is the velocity of the wakefield ($v_{\rm w} < c$), the total phase retardation $\delta \phi$ is thus limited by $d_\gamma$.
With current laser and plasma parameters, one obtains $d_{\gamma, \rm optimal} \simeq 27~ \mu{\rm m}$.
When $d_\gamma < d_{\gamma, \rm optimal}$, $L < L_{\rm max}$, and $\delta \phi$ and $\mathcal{A}$ will be smaller.
When $d_\gamma > d_{\gamma, \rm optimal}$, the probe $\gamma$-photons are located near the tail of the bubble. Thus the average wakefield experienced by them will be weaker, and the VB signal smaller [Fig.~\ref{fig4}(a)].
The  quality of the probe beam may also influence the VB signal. The angular spread $\Delta\theta$ will reduce the VB signal [Fig.~\ref{fig4}(b)], however, even at $\Delta\theta \simeq 20$~mrad, one still obtains $\mathcal{A}\simeq 10^{-5}$.
In the case of the beam-like energy distribution around the central energy, e.g. \cite{Weller2015,Li_2020_Polarized}, the energy spreading does not disturb much the VB signal [Fig.~\ref{fig4}(c)], while in the case of a Maxwellian distribution, $\mathcal{A}$ is  proportional to the ``temperature"  of the probe beam [Fig.~\ref{fig4}(d)]. The impact of beam radii $r_t$ and $r_l$ on the signatures of VB are discussed in Sec. V of \cite{supplemental}.

We note finally that the bubble wakefield can be created also by ultrarelativistic electron or proton beams \cite{Hidding2019,awake}. However, in the first case NCS is significant and should be spatially separated from the probe photons. In the second case, the large wakefield requires larger proton densities presently not available with conventional schemes \cite{Shen2011,Zheng2012}.

In conclusion, we put forward a competitive VB detection method based on a plasma wakefield and using moderate laser intensities, which can provide a VB signal $\sim  10^{-5}$ for MeV-level probe photons and $\sim  10^{-3}$-$10^{-2}$ for GeV level within several millimeters of the interaction length. The noise because of plasma electrons is shown to be reducible the signal level. The method is robust with respect to laser and plasma parameters and can be realized with currently feasible laser facilities.
Furthermore, the method has the potential to
apply to the search for axion-like particles, as the VB could be
mediated via weakly interacting slim particles (WISPs) rather than
electron-positrons.

{\it Acknowledgment:} The work is supported by the National Natural Science Foundation of China (Grants Nos. 12022506, 11874295, 11875219, 11905169, 12005305), the National Key R\&D Program of China No. 2019YFA0404900, Open Foundation of Key Laboratory of High Power Laser and Physics, Chinese Academy of Sciences (SGKF202101), the China Postdoctoral Science Foundation (Grant No. 2020M683447), the Open Fund of the State Key Laboratory of High Field Laser Physics (Shanghai Institute of Optics and Fine Mechanics) and the Fundamental Research Funds for the Central Universities.

\bibliography{vblib}

\begin{thebibliography}{72}%
\makeatletter
\providecommand \@ifxundefined [1]{%
 \@ifx{#1\undefined}
}%
\providecommand \@ifnum [1]{%
 \ifnum #1\expandafter \@firstoftwo
 \else \expandafter \@secondoftwo
 \fi
}%
\providecommand \@ifx [1]{%
 \ifx #1\expandafter \@firstoftwo
 \else \expandafter \@secondoftwo
 \fi
}%
\providecommand \natexlab [1]{#1}%
\providecommand \enquote  [1]{``#1''}%
\providecommand \bibnamefont  [1]{#1}%
\providecommand \bibfnamefont [1]{#1}%
\providecommand \citenamefont [1]{#1}%
\providecommand \href@noop [0]{\@secondoftwo}%
\providecommand \href [0]{\begingroup \@sanitize@url \@href}%
\providecommand \@href[1]{\@@startlink{#1}\@@href}%
\providecommand \@@href[1]{\endgroup#1\@@endlink}%
\providecommand \@sanitize@url [0]{\catcode `\\12\catcode `\$12\catcode `\&12\catcode `\#12\catcode `\^12\catcode `\_12\catcode `\%12\relax}%
\providecommand \@@startlink[1]{}%
\providecommand \@@endlink[0]{}%
\providecommand \url  [0]{\begingroup\@sanitize@url \@url }%
\providecommand \@url [1]{\endgroup\@href {#1}{\urlprefix }}%
\providecommand \urlprefix  [0]{URL }%
\providecommand \Eprint [0]{\href }%
\providecommand \doibase [0]{https://doi.org/}%
\providecommand \selectlanguage [0]{\@gobble}%
\providecommand \bibinfo  [0]{\@secondoftwo}%
\providecommand \bibfield  [0]{\@secondoftwo}%
\providecommand \translation [1]{[#1]}%
\providecommand \BibitemOpen [0]{}%
\providecommand \bibitemStop [0]{}%
\providecommand \bibitemNoStop [0]{.\EOS\space}%
\providecommand \EOS [0]{\spacefactor3000\relax}%
\providecommand \BibitemShut  [1]{\csname bibitem#1\endcsname}%
\let\auto@bib@innerbib\@empty
\bibitem [{\citenamefont {Halpern}(1933)}]{Halpern_1933}%
  \BibitemOpen
  \bibfield  {author} {\bibinfo {author} {\bibfnamefont {O.}~\bibnamefont {Halpern}},\ }\bibfield  {title} {\bibinfo {title} {Scattering processes produced by electrons in negative energy states},\ }\href {https://doi.org/10.1103/PhysRev.44.855.2} {\bibfield  {journal} {\bibinfo  {journal} {Phys. Rev.}\ }\textbf {\bibinfo {volume} {44}},\ \bibinfo {pages} {855} (\bibinfo {year} {1933})}\BibitemShut {NoStop}%
\bibitem [{\citenamefont {Euler}\ and\ \citenamefont {Kockel}(1935)}]{Euler_1935}%
  \BibitemOpen
  \bibfield  {author} {\bibinfo {author} {\bibfnamefont {H.}~\bibnamefont {Euler}}\ and\ \bibinfo {author} {\bibfnamefont {B.}~\bibnamefont {Kockel}},\ }\bibfield  {title} {\bibinfo {title} {{The scattering of light by light in {D}irac's theory}},\ }\href {https://doi.org/10.1007/BF01493898} {\bibfield  {journal} {\bibinfo  {journal} {Naturwiss.}\ }\textbf {\bibinfo {volume} {23}},\ \bibinfo {pages} {246} (\bibinfo {year} {1935})}\BibitemShut {NoStop}%
\bibitem [{\citenamefont {Schwinger}(1949)}]{Schwinger_1949_Quantum}%
  \BibitemOpen
  \bibfield  {author} {\bibinfo {author} {\bibfnamefont {J.}~\bibnamefont {Schwinger}},\ }\bibfield  {title} {\bibinfo {title} {Quantum {E}lectrodynamics. {II}. {V}acuum {P}olarization and {S}elf-{E}nergy},\ }\href {https://doi.org/10.1103/physrev.75.651} {\bibfield  {journal} {\bibinfo  {journal} {Phys. Rev.}\ }\textbf {\bibinfo {volume} {75}},\ \bibinfo {pages} {651} (\bibinfo {year} {1949})}\BibitemShut {NoStop}%
\bibitem [{\citenamefont {Schwinger}(1951)}]{Schwinger_1951_Gauge}%
  \BibitemOpen
  \bibfield  {author} {\bibinfo {author} {\bibfnamefont {J.}~\bibnamefont {Schwinger}},\ }\bibfield  {title} {\bibinfo {title} {On {G}auge {I}nvariance and {V}acuum {P}olarization},\ }\href {https://doi.org/10.1103/physrev.82.664} {\bibfield  {journal} {\bibinfo  {journal} {Phys. Rev.}\ }\textbf {\bibinfo {volume} {82}},\ \bibinfo {pages} {664} (\bibinfo {year} {1951})}\BibitemShut {NoStop}%
\bibitem [{\citenamefont {Klein}\ and\ \citenamefont {Nigam}(1964)}]{Klein_1964_Birefringence}%
  \BibitemOpen
  \bibfield  {author} {\bibinfo {author} {\bibfnamefont {J.~J.}\ \bibnamefont {Klein}}\ and\ \bibinfo {author} {\bibfnamefont {B.~P.}\ \bibnamefont {Nigam}},\ }\bibfield  {title} {\bibinfo {title} {Birefringence of the {V}acuum},\ }\href {https://doi.org/10.1103/physrev.135.b1279} {\bibfield  {journal} {\bibinfo  {journal} {Phys. Rev.}\ }\textbf {\bibinfo {volume} {135}},\ \bibinfo {pages} {B1279} (\bibinfo {year} {1964})}\BibitemShut {NoStop}%
\bibitem [{\citenamefont {Brezin}\ and\ \citenamefont {Itzykson}(1971)}]{Brezin_1971_Polarization}%
  \BibitemOpen
  \bibfield  {author} {\bibinfo {author} {\bibfnamefont {E.}~\bibnamefont {Brezin}}\ and\ \bibinfo {author} {\bibfnamefont {C.}~\bibnamefont {Itzykson}},\ }\bibfield  {title} {\bibinfo {title} {Polarization {P}henomena in {V}acuum {N}onlinear {E}lectrodynamics},\ }\href {https://doi.org/10.1103/physrevd.3.618} {\bibfield  {journal} {\bibinfo  {journal} {Phys. Rev. D}\ }\textbf {\bibinfo {volume} {3}},\ \bibinfo {pages} {618} (\bibinfo {year} {1971})}\BibitemShut {NoStop}%
\bibitem [{\citenamefont {Li}\ \emph {et~al.}(2015)\citenamefont {Li}, \citenamefont {Lu},\ and\ \citenamefont {Xie}}]{XiePRD}%
  \BibitemOpen
  \bibfield  {author} {\bibinfo {author} {\bibfnamefont {Z.~L.}\ \bibnamefont {Li}}, \bibinfo {author} {\bibfnamefont {D.}~\bibnamefont {Lu}},\ and\ \bibinfo {author} {\bibfnamefont {B.~S.}\ \bibnamefont {Xie}},\ }\bibfield  {title} {\bibinfo {title} {Effects of electric field polarizations on pair production},\ }\href {https://doi.org/10.1103/PhysRevD.92.085001} {\bibfield  {journal} {\bibinfo  {journal} {Phys. Rev. D}\ }\textbf {\bibinfo {volume} {92}},\ \bibinfo {pages} {085001} (\bibinfo {year} {2015})}\BibitemShut {NoStop}%
\bibitem [{\citenamefont {Dittrich}(2000)}]{dittrich2000probing}%
  \BibitemOpen
  \bibfield  {author} {\bibinfo {author} {\bibfnamefont {W.}~\bibnamefont {Dittrich}},\ }\href {https://doi.org/10.1007/3-540-45585-x} {\emph {\bibinfo {title} {Probing the quantum vacuum : pertubative effective action approach in quantum electrodynamics and its application}}}\ (\bibinfo  {publisher} {Springer},\ \bibinfo {address} {Berlin New York},\ \bibinfo {year} {2000})\BibitemShut {NoStop}%
\bibitem [{\citenamefont {Marklund}\ and\ \citenamefont {Shukla}(2006)}]{Marklund2006}%
  \BibitemOpen
  \bibfield  {author} {\bibinfo {author} {\bibfnamefont {M.}~\bibnamefont {Marklund}}\ and\ \bibinfo {author} {\bibfnamefont {P.~K.}\ \bibnamefont {Shukla}},\ }\bibfield  {title} {\bibinfo {title} {Nonlinear collective effects in photon-photon and photon-plasma interactions},\ }\href {https://doi.org/10.1103/revmodphys.78.591} {\bibfield  {journal} {\bibinfo  {journal} {Rev. Mod. Phys.}\ }\textbf {\bibinfo {volume} {78}},\ \bibinfo {pages} {591} (\bibinfo {year} {2006})}\BibitemShut {NoStop}%
\bibitem [{\citenamefont {Piazza}\ \emph {et~al.}(2012)\citenamefont {Piazza}, \citenamefont {Müller}, \citenamefont {Hatsagortsyan},\ and\ \citenamefont {Keitel}}]{Piazza2012}%
  \BibitemOpen
  \bibfield  {author} {\bibinfo {author} {\bibfnamefont {A.~D.}\ \bibnamefont {Piazza}}, \bibinfo {author} {\bibfnamefont {C.}~\bibnamefont {Müller}}, \bibinfo {author} {\bibfnamefont {K.~Z.}\ \bibnamefont {Hatsagortsyan}},\ and\ \bibinfo {author} {\bibfnamefont {C.~H.}\ \bibnamefont {Keitel}},\ }\bibfield  {title} {\bibinfo {title} {Extremely high-intensity laser interactions with fundamental quantum systems},\ }\href {https://doi.org/10.1103/revmodphys.84.1177} {\bibfield  {journal} {\bibinfo  {journal} {Rev. Mod. Phys.}\ }\textbf {\bibinfo {volume} {84}},\ \bibinfo {pages} {1177} (\bibinfo {year} {2012})}\BibitemShut {NoStop}%
\bibitem [{\citenamefont {Karbstein}(2013)}]{Karbstein_2013_Photon}%
  \BibitemOpen
  \bibfield  {author} {\bibinfo {author} {\bibfnamefont {F.}~\bibnamefont {Karbstein}},\ }\bibfield  {title} {\bibinfo {title} {Photon polarization tensor in a homogeneous magnetic or electric field},\ }\href {https://doi.org/10.1103/physrevd.88.085033} {\bibfield  {journal} {\bibinfo  {journal} {Phys. Rev. D}\ }\textbf {\bibinfo {volume} {88}},\ \bibinfo {pages} {085033} (\bibinfo {year} {2013})}\BibitemShut {NoStop}%
\bibitem [{\citenamefont {Gonoskov}\ \emph {et~al.}()\citenamefont {Gonoskov}, \citenamefont {Blackburn}, \citenamefont {Marklund},\ and\ \citenamefont {Bulanov}}]{Gonoskov2021}%
  \BibitemOpen
  \bibfield  {author} {\bibinfo {author} {\bibfnamefont {A.}~\bibnamefont {Gonoskov}}, \bibinfo {author} {\bibfnamefont {T.~G.}\ \bibnamefont {Blackburn}}, \bibinfo {author} {\bibfnamefont {M.}~\bibnamefont {Marklund}},\ and\ \bibinfo {author} {\bibfnamefont {S.~S.}\ \bibnamefont {Bulanov}},\ }\href {https://doi.org/10.3103/s1541308x21010039} {\bibinfo {title} {Charged particle motion and radiation in strong electromagnetic fields}},\ \Eprint {https://arxiv.org/abs/2107.02161} {arXiv:2107.02161} \BibitemShut {NoStop}%
\bibitem [{\citenamefont {Ejlli}\ \emph {et~al.}(2020)\citenamefont {Ejlli}, \citenamefont {Valle}, \citenamefont {Gastaldi}, \citenamefont {Messineo}, \citenamefont {Pengo}, \citenamefont {Ruoso},\ and\ \citenamefont {Zavattini}}]{Ejlli2020}%
  \BibitemOpen
  \bibfield  {author} {\bibinfo {author} {\bibfnamefont {A.}~\bibnamefont {Ejlli}}, \bibinfo {author} {\bibfnamefont {F.~D.}\ \bibnamefont {Valle}}, \bibinfo {author} {\bibfnamefont {U.}~\bibnamefont {Gastaldi}}, \bibinfo {author} {\bibfnamefont {G.}~\bibnamefont {Messineo}}, \bibinfo {author} {\bibfnamefont {R.}~\bibnamefont {Pengo}}, \bibinfo {author} {\bibfnamefont {G.}~\bibnamefont {Ruoso}},\ and\ \bibinfo {author} {\bibfnamefont {G.}~\bibnamefont {Zavattini}},\ }\bibfield  {title} {\bibinfo {title} {The {PVLAS} experiment: A 25 year effort to measure vacuum magnetic birefringence},\ }\href {https://doi.org/10.1016/j.physrep.2020.06.001} {\bibfield  {journal} {\bibinfo  {journal} {Phys. Rep.}\ }\textbf {\bibinfo {volume} {871}},\ \bibinfo {pages} {1} (\bibinfo {year} {2020})}\BibitemShut {NoStop}%
\bibitem [{\citenamefont {Cameron}\ \emph {et~al.}(1993)\citenamefont {Cameron}, \citenamefont {Cantatore}, \citenamefont {Melissinos}, \citenamefont {Ruoso}, \citenamefont {Semertzidis}, \citenamefont {Halama}, \citenamefont {Lazarus}, \citenamefont {Prodell}, \citenamefont {Nezrick}, \citenamefont {Rizzo},\ and\ \citenamefont {Zavattini}}]{Cameron1993}%
  \BibitemOpen
  \bibfield  {author} {\bibinfo {author} {\bibfnamefont {R.}~\bibnamefont {Cameron}}, \bibinfo {author} {\bibfnamefont {G.}~\bibnamefont {Cantatore}}, \bibinfo {author} {\bibfnamefont {A.~C.}\ \bibnamefont {Melissinos}}, \bibinfo {author} {\bibfnamefont {G.}~\bibnamefont {Ruoso}}, \bibinfo {author} {\bibfnamefont {Y.}~\bibnamefont {Semertzidis}}, \bibinfo {author} {\bibfnamefont {H.~J.}\ \bibnamefont {Halama}}, \bibinfo {author} {\bibfnamefont {D.~M.}\ \bibnamefont {Lazarus}}, \bibinfo {author} {\bibfnamefont {A.~G.}\ \bibnamefont {Prodell}}, \bibinfo {author} {\bibfnamefont {F.}~\bibnamefont {Nezrick}}, \bibinfo {author} {\bibfnamefont {C.}~\bibnamefont {Rizzo}},\ and\ \bibinfo {author} {\bibfnamefont {E.}~\bibnamefont {Zavattini}},\ }\bibfield  {title} {\bibinfo {title} {Search for nearly massless, weakly coupled particles by optical techniques},\ }\href {https://doi.org/10.1103/physrevd.47.3707} {\bibfield  {journal} {\bibinfo  {journal} {Phys. Rev. D}\ }\textbf {\bibinfo {volume} {47}},\ \bibinfo {pages} {3707} (\bibinfo {year} {1993})}\BibitemShut {NoStop}%
\bibitem [{\citenamefont {Cad{\`{e}}ne}\ \emph {et~al.}(2014)\citenamefont {Cad{\`{e}}ne}, \citenamefont {Berceau}, \citenamefont {Fouch{\'{e}}}, \citenamefont {Battesti},\ and\ \citenamefont {Rizzo}}]{Cadene2014}%
  \BibitemOpen
  \bibfield  {author} {\bibinfo {author} {\bibfnamefont {A.}~\bibnamefont {Cad{\`{e}}ne}}, \bibinfo {author} {\bibfnamefont {P.}~\bibnamefont {Berceau}}, \bibinfo {author} {\bibfnamefont {M.}~\bibnamefont {Fouch{\'{e}}}}, \bibinfo {author} {\bibfnamefont {R.}~\bibnamefont {Battesti}},\ and\ \bibinfo {author} {\bibfnamefont {C.}~\bibnamefont {Rizzo}},\ }\bibfield  {title} {\bibinfo {title} {{Vacuum magnetic linear birefringence using pulsed fields: status of the {BMV} experiment}},\ }\href {https://doi.org/10.1140/epjd/e2013-40725-9} {\bibfield  {journal} {\bibinfo  {journal} {Eur. Phys. J. D}\ }\textbf {\bibinfo {volume} {68}},\ \bibinfo {pages} {16} (\bibinfo {year} {2014})}\BibitemShut {NoStop}%
\bibitem [{\citenamefont {Cantatore}\ \emph {et~al.}(1991)\citenamefont {Cantatore}, \citenamefont {Valle}, \citenamefont {Milotti}, \citenamefont {Dabrowski},\ and\ \citenamefont {Rizzo}}]{Cantatore_1991_Proposed}%
  \BibitemOpen
  \bibfield  {author} {\bibinfo {author} {\bibfnamefont {G.}~\bibnamefont {Cantatore}}, \bibinfo {author} {\bibfnamefont {F.~D.}\ \bibnamefont {Valle}}, \bibinfo {author} {\bibfnamefont {E.}~\bibnamefont {Milotti}}, \bibinfo {author} {\bibfnamefont {L.}~\bibnamefont {Dabrowski}},\ and\ \bibinfo {author} {\bibfnamefont {C.}~\bibnamefont {Rizzo}},\ }\bibfield  {title} {\bibinfo {title} {Proposed measurement of the vacuum birefringence induced by a magnetic field on high energy photons},\ }\href {https://doi.org/10.1016/0370-2693(91)90077-4} {\bibfield  {journal} {\bibinfo  {journal} {Phys. Lett. B}\ }\textbf {\bibinfo {volume} {265}},\ \bibinfo {pages} {418} (\bibinfo {year} {1991})}\BibitemShut {NoStop}%
\bibitem [{\citenamefont {Wistisen}\ and\ \citenamefont {Uggerh{\o}j}(2013)}]{Wistisen_2013_Vacuum}%
  \BibitemOpen
  \bibfield  {author} {\bibinfo {author} {\bibfnamefont {T.~N.}\ \bibnamefont {Wistisen}}\ and\ \bibinfo {author} {\bibfnamefont {U.~I.}\ \bibnamefont {Uggerh{\o}j}},\ }\bibfield  {title} {\bibinfo {title} {Vacuum birefringence by {C}ompton backscattering through a strong field},\ }\href {https://doi.org/10.1103/physrevd.88.053009} {\bibfield  {journal} {\bibinfo  {journal} {Phys. Rev. D}\ }\textbf {\bibinfo {volume} {88}},\ \bibinfo {pages} {053009} (\bibinfo {year} {2013})}\BibitemShut {NoStop}%
\bibitem [{\citenamefont {Gales}\ \emph {et~al.}(2018)\citenamefont {Gales}, \citenamefont {Tanaka}, \citenamefont {Balabanski}, \citenamefont {Negoita}, \citenamefont {Stutman}, \citenamefont {Tesileanu}, \citenamefont {Ur}, \citenamefont {Ursescu}, \citenamefont {Andrei}, \citenamefont {Ataman}, \citenamefont {Cernaianu}, \citenamefont {D'Alessi}, \citenamefont {Dancus}, \citenamefont {Diaconescu}, \citenamefont {Djourelov}, \citenamefont {Filipescu}, \citenamefont {Ghenuche}, \citenamefont {Ghita}, \citenamefont {Matei}, \citenamefont {Seto}, \citenamefont {Zeng},\ and\ \citenamefont {Zamfir}}]{Gales_2018_extreme}%
  \BibitemOpen
  \bibfield  {author} {\bibinfo {author} {\bibfnamefont {S.}~\bibnamefont {Gales}}, \bibinfo {author} {\bibfnamefont {K.~A.}\ \bibnamefont {Tanaka}}, \bibinfo {author} {\bibfnamefont {D.~L.}\ \bibnamefont {Balabanski}}, \bibinfo {author} {\bibfnamefont {F.}~\bibnamefont {Negoita}}, \bibinfo {author} {\bibfnamefont {D.}~\bibnamefont {Stutman}}, \bibinfo {author} {\bibfnamefont {O.}~\bibnamefont {Tesileanu}}, \bibinfo {author} {\bibfnamefont {C.~A.}\ \bibnamefont {Ur}}, \bibinfo {author} {\bibfnamefont {D.}~\bibnamefont {Ursescu}}, \bibinfo {author} {\bibfnamefont {I.}~\bibnamefont {Andrei}}, \bibinfo {author} {\bibfnamefont {S.}~\bibnamefont {Ataman}}, \bibinfo {author} {\bibfnamefont {M.~O.}\ \bibnamefont {Cernaianu}}, \bibinfo {author} {\bibfnamefont {L.}~\bibnamefont {D'Alessi}}, \bibinfo {author} {\bibfnamefont {I.}~\bibnamefont {Dancus}}, \bibinfo {author} {\bibfnamefont {B.}~\bibnamefont {Diaconescu}}, \bibinfo {author} {\bibfnamefont {N.}~\bibnamefont {Djourelov}}, \bibinfo {author} {\bibfnamefont {D.}~\bibnamefont {Filipescu}}, \bibinfo {author} {\bibfnamefont {P.}~\bibnamefont {Ghenuche}}, \bibinfo {author} {\bibfnamefont {D.~G.}\ \bibnamefont {Ghita}}, \bibinfo {author} {\bibfnamefont {C.}~\bibnamefont {Matei}}, \bibinfo {author} {\bibfnamefont {K.}~\bibnamefont {Seto}}, \bibinfo {author} {\bibfnamefont {M.}~\bibnamefont {Zeng}},\ and\ \bibinfo {author} {\bibfnamefont {N.~V.}\ \bibnamefont {Zamfir}},\ }\bibfield  {title} {\bibinfo {title} {The extreme light infrastructure—nuclear physics ({ELI}-{NP}) facility: new horizons in physics with 10 {PW} ultra-intense lasers and 20 {MeV} brilliant gamma beams},\ }\href {https://doi.org/10.1088/1361-6633/aacfe8} {\bibfield  {journal} {\bibinfo  {journal} {Rep. Progr. Phys.}\ }\textbf {\bibinfo {volume} {81}},\ \bibinfo {pages} {094301} (\bibinfo {year} {2018})}\BibitemShut {NoStop}%
\bibitem [{\citenamefont {Danson}\ \emph {et~al.}(2019)\citenamefont {Danson}, \citenamefont {Haefner}, \citenamefont {Bromage}, \citenamefont {Butcher}, \citenamefont {Chanteloup}, \citenamefont {Chowdhury}, \citenamefont {Galvanauskas}, \citenamefont {Gizzi}, \citenamefont {Hein}, \citenamefont {Hillier}, \citenamefont {Hopps}, \citenamefont {Kato}, \citenamefont {Khazanov}, \citenamefont {Kodama}, \citenamefont {Korn}, \citenamefont {Li}, \citenamefont {Li}, \citenamefont {Limpert}, \citenamefont {Ma}, \citenamefont {Nam}, \citenamefont {Neely}, \citenamefont {Papadopoulos}, \citenamefont {Penman}, \citenamefont {Qian}, \citenamefont {Rocca}, \citenamefont {Shaykin}, \citenamefont {Siders}, \citenamefont {Spindloe}, \citenamefont {Szatm{\'{a}}ri}, \citenamefont {Trines}, \citenamefont {Zhu}, \citenamefont {Zhu},\ and\ \citenamefont {Zuegel}}]{Danson_2019_Petawatt}%
  \BibitemOpen
  \bibfield  {author} {\bibinfo {author} {\bibfnamefont {C.~N.}\ \bibnamefont {Danson}}, \bibinfo {author} {\bibfnamefont {C.}~\bibnamefont {Haefner}}, \bibinfo {author} {\bibfnamefont {J.}~\bibnamefont {Bromage}}, \bibinfo {author} {\bibfnamefont {T.}~\bibnamefont {Butcher}}, \bibinfo {author} {\bibfnamefont {J.-C.~F.}\ \bibnamefont {Chanteloup}}, \bibinfo {author} {\bibfnamefont {E.~A.}\ \bibnamefont {Chowdhury}}, \bibinfo {author} {\bibfnamefont {A.}~\bibnamefont {Galvanauskas}}, \bibinfo {author} {\bibfnamefont {L.~A.}\ \bibnamefont {Gizzi}}, \bibinfo {author} {\bibfnamefont {J.}~\bibnamefont {Hein}}, \bibinfo {author} {\bibfnamefont {D.~I.}\ \bibnamefont {Hillier}}, \bibinfo {author} {\bibfnamefont {N.~W.}\ \bibnamefont {Hopps}}, \bibinfo {author} {\bibfnamefont {Y.}~\bibnamefont {Kato}}, \bibinfo {author} {\bibfnamefont {E.~A.}\ \bibnamefont {Khazanov}}, \bibinfo {author} {\bibfnamefont {R.}~\bibnamefont {Kodama}}, \bibinfo {author} {\bibfnamefont {G.}~\bibnamefont {Korn}}, \bibinfo {author} {\bibfnamefont {R.}~\bibnamefont {Li}}, \bibinfo {author} {\bibfnamefont {Y.}~\bibnamefont {Li}}, \bibinfo {author} {\bibfnamefont {J.}~\bibnamefont {Limpert}}, \bibinfo {author} {\bibfnamefont {J.}~\bibnamefont {Ma}}, \bibinfo {author} {\bibfnamefont {C.~H.}\ \bibnamefont {Nam}}, \bibinfo {author} {\bibfnamefont {D.}~\bibnamefont {Neely}}, \bibinfo {author} {\bibfnamefont {D.}~\bibnamefont {Papadopoulos}}, \bibinfo {author} {\bibfnamefont {R.~R.}\ \bibnamefont {Penman}}, \bibinfo {author} {\bibfnamefont {L.}~\bibnamefont {Qian}}, \bibinfo {author} {\bibfnamefont {J.~J.}\ \bibnamefont {Rocca}}, \bibinfo {author} {\bibfnamefont {A.~A.}\ \bibnamefont {Shaykin}}, \bibinfo {author} {\bibfnamefont {C.~W.}\ \bibnamefont {Siders}}, \bibinfo {author} {\bibfnamefont {C.}~\bibnamefont {Spindloe}}, \bibinfo {author} {\bibfnamefont {S.}~\bibnamefont {Szatm{\'{a}}ri}}, \bibinfo {author} {\bibfnamefont {R.~M. G.~M.}\ \bibnamefont {Trines}}, \bibinfo {author} {\bibfnamefont {J.}~\bibnamefont {Zhu}}, \bibinfo {author} {\bibfnamefont {P.}~\bibnamefont {Zhu}},\ and\ \bibinfo {author} {\bibfnamefont {J.~D.}\ \bibnamefont {Zuegel}},\ }\bibfield  {title} {\bibinfo {title} {Petawatt and exawatt class lasers worldwide},\ }\href {https://doi.org/10.1017/hpl.2019.36} {\bibfield  {journal} {\bibinfo  {journal} {High Power Laser Sci. Eng.}\ }\textbf {\bibinfo {volume} {7}},\ \bibinfo {pages} {e54} (\bibinfo {year} {2019})}\BibitemShut {NoStop}%
\bibitem [{\citenamefont {Yoon}\ \emph {et~al.}(2019)\citenamefont {Yoon}, \citenamefont {Jeon}, \citenamefont {Shin}, \citenamefont {Lee}, \citenamefont {Lee}, \citenamefont {Choi}, \citenamefont {Kim}, \citenamefont {Sung},\ and\ \citenamefont {Nam}}]{Yoon_2019_Achieving}%
  \BibitemOpen
  \bibfield  {author} {\bibinfo {author} {\bibfnamefont {J.~W.}\ \bibnamefont {Yoon}}, \bibinfo {author} {\bibfnamefont {C.}~\bibnamefont {Jeon}}, \bibinfo {author} {\bibfnamefont {J.}~\bibnamefont {Shin}}, \bibinfo {author} {\bibfnamefont {S.~K.}\ \bibnamefont {Lee}}, \bibinfo {author} {\bibfnamefont {H.~W.}\ \bibnamefont {Lee}}, \bibinfo {author} {\bibfnamefont {I.~W.}\ \bibnamefont {Choi}}, \bibinfo {author} {\bibfnamefont {H.~T.}\ \bibnamefont {Kim}}, \bibinfo {author} {\bibfnamefont {J.~H.}\ \bibnamefont {Sung}},\ and\ \bibinfo {author} {\bibfnamefont {C.~H.}\ \bibnamefont {Nam}},\ }\bibfield  {title} {\bibinfo {title} {{Achieving the laser intensity of $5.5\times10^{22} {\rm W/cm}^2$ with a wavefront-corrected multi-{PW} laser}},\ }\href {https://doi.org/10.1364/oe.27.020412} {\bibfield  {journal} {\bibinfo  {journal} {Opt. Express}\ }\textbf {\bibinfo {volume} {27}},\ \bibinfo {pages} {20412} (\bibinfo {year} {2019})}\BibitemShut {NoStop}%
\bibitem [{\citenamefont {Yoon}\ \emph {et~al.}(2021)\citenamefont {Yoon}, \citenamefont {Kim}, \citenamefont {Choi}, \citenamefont {Sung}, \citenamefont {Lee}, \citenamefont {Lee},\ and\ \citenamefont {Nam}}]{Yoon_2021}%
  \BibitemOpen
  \bibfield  {author} {\bibinfo {author} {\bibfnamefont {J.~W.}\ \bibnamefont {Yoon}}, \bibinfo {author} {\bibfnamefont {Y.~G.}\ \bibnamefont {Kim}}, \bibinfo {author} {\bibfnamefont {I.~W.}\ \bibnamefont {Choi}}, \bibinfo {author} {\bibfnamefont {J.~H.}\ \bibnamefont {Sung}}, \bibinfo {author} {\bibfnamefont {H.~W.}\ \bibnamefont {Lee}}, \bibinfo {author} {\bibfnamefont {S.~K.}\ \bibnamefont {Lee}},\ and\ \bibinfo {author} {\bibfnamefont {C.~H.}\ \bibnamefont {Nam}},\ }\bibfield  {title} {\bibinfo {title} {Realization of laser intensity over $10^{23}$ {W}/cm$^2$},\ }\href {https://doi.org/10.1364/OPTICA.420520} {\bibfield  {journal} {\bibinfo  {journal} {Optica}\ }\textbf {\bibinfo {volume} {8}},\ \bibinfo {pages} {630} (\bibinfo {year} {2021})}\BibitemShut {NoStop}%
\bibitem [{\citenamefont {Di~Piazza}\ \emph {et~al.}(2006)\citenamefont {Di~Piazza}, \citenamefont {Hatsagortsyan},\ and\ \citenamefont {Keitel}}]{DiPiazza_2006}%
  \BibitemOpen
  \bibfield  {author} {\bibinfo {author} {\bibfnamefont {A.}~\bibnamefont {Di~Piazza}}, \bibinfo {author} {\bibfnamefont {K.~Z.}\ \bibnamefont {Hatsagortsyan}},\ and\ \bibinfo {author} {\bibfnamefont {C.~H.}\ \bibnamefont {Keitel}},\ }\bibfield  {title} {\bibinfo {title} {Light diffraction by a strong standing electromagnetic wave},\ }\href {https://doi.org/10.1103/PhysRevLett.97.083603} {\bibfield  {journal} {\bibinfo  {journal} {Phys. Rev. Lett.}\ }\textbf {\bibinfo {volume} {97}},\ \bibinfo {pages} {083603} (\bibinfo {year} {2006})}\BibitemShut {NoStop}%
\bibitem [{\citenamefont {Heinzl}\ \emph {et~al.}(2006)\citenamefont {Heinzl}, \citenamefont {Liesfeld}, \citenamefont {Amthor}, \citenamefont {Schwoerer}, \citenamefont {Sauerbrey},\ and\ \citenamefont {Wipf}}]{Heinzl_2006_observation}%
  \BibitemOpen
  \bibfield  {author} {\bibinfo {author} {\bibfnamefont {T.}~\bibnamefont {Heinzl}}, \bibinfo {author} {\bibfnamefont {B.}~\bibnamefont {Liesfeld}}, \bibinfo {author} {\bibfnamefont {K.-U.}\ \bibnamefont {Amthor}}, \bibinfo {author} {\bibfnamefont {H.}~\bibnamefont {Schwoerer}}, \bibinfo {author} {\bibfnamefont {R.}~\bibnamefont {Sauerbrey}},\ and\ \bibinfo {author} {\bibfnamefont {A.}~\bibnamefont {Wipf}},\ }\bibfield  {title} {\bibinfo {title} {On the observation of vacuum birefringence},\ }\href {https://doi.org/10.1016/j.optcom.2006.06.053} {\bibfield  {journal} {\bibinfo  {journal} {Opt. Commun.}\ }\textbf {\bibinfo {volume} {267}},\ \bibinfo {pages} {318} (\bibinfo {year} {2006})}\BibitemShut {NoStop}%
\bibitem [{\citenamefont {Di~Piazza}\ \emph {et~al.}(2007)\citenamefont {Di~Piazza}, \citenamefont {Hatsagortsyan},\ and\ \citenamefont {Keitel}}]{Piazza2007Enhancement}%
  \BibitemOpen
  \bibfield  {author} {\bibinfo {author} {\bibfnamefont {A.}~\bibnamefont {Di~Piazza}}, \bibinfo {author} {\bibfnamefont {K.~Z.}\ \bibnamefont {Hatsagortsyan}},\ and\ \bibinfo {author} {\bibfnamefont {C.~H.}\ \bibnamefont {Keitel}},\ }\bibfield  {title} {\bibinfo {title} {{Enhancement of vacuum polarization effects in a plasma}},\ }\href {https://doi.org/10.1063/1.2646541} {\bibfield  {journal} {\bibinfo  {journal} {Physics of Plasmas}\ }\textbf {\bibinfo {volume} {14}},\ \bibinfo {pages} {032102} (\bibinfo {year} {2007})}\BibitemShut {NoStop}%
\bibitem [{\citenamefont {Klar}(2020)}]{Klar2020}%
  \BibitemOpen
  \bibfield  {author} {\bibinfo {author} {\bibfnamefont {L.}~\bibnamefont {Klar}},\ }\bibfield  {title} {\bibinfo {title} {{Detectable Optical Signatures of {QED} Vacuum Nonlinearities Using High-Intensity Laser Fields}},\ }\href {https://doi.org/10.3390/particles3010018} {\bibfield  {journal} {\bibinfo  {journal} {Particles}\ }\textbf {\bibinfo {volume} {3}},\ \bibinfo {pages} {223} (\bibinfo {year} {2020})}\BibitemShut {NoStop}%
\bibitem [{\citenamefont {Schlenvoigt}\ \emph {et~al.}(2016)\citenamefont {Schlenvoigt}, \citenamefont {Heinzl}, \citenamefont {Schramm}, \citenamefont {Cowan},\ and\ \citenamefont {Sauerbrey}}]{Schlenvoigt_2016_Detecting}%
  \BibitemOpen
  \bibfield  {author} {\bibinfo {author} {\bibfnamefont {H.-P.}\ \bibnamefont {Schlenvoigt}}, \bibinfo {author} {\bibfnamefont {T.}~\bibnamefont {Heinzl}}, \bibinfo {author} {\bibfnamefont {U.}~\bibnamefont {Schramm}}, \bibinfo {author} {\bibfnamefont {T.~E.}\ \bibnamefont {Cowan}},\ and\ \bibinfo {author} {\bibfnamefont {R.}~\bibnamefont {Sauerbrey}},\ }\bibfield  {title} {\bibinfo {title} {Detecting vacuum birefringence with x-ray free electron lasers and high-power optical lasers: a feasibility study},\ }\href {https://doi.org/10.1088/0031-8949/91/2/023010} {\bibfield  {journal} {\bibinfo  {journal} {Phys. Scripta}\ }\textbf {\bibinfo {volume} {91}},\ \bibinfo {pages} {023010} (\bibinfo {year} {2016})}\BibitemShut {NoStop}%
\bibitem [{\citenamefont {Karbstein}\ and\ \citenamefont {Sundqvist}(2016)}]{Karbstein_2016lby}%
  \BibitemOpen
  \bibfield  {author} {\bibinfo {author} {\bibfnamefont {F.}~\bibnamefont {Karbstein}}\ and\ \bibinfo {author} {\bibfnamefont {C.}~\bibnamefont {Sundqvist}},\ }\bibfield  {title} {\bibinfo {title} {{Probing vacuum birefringence using x-ray free electron and optical high-intensity lasers}},\ }\href {https://doi.org/10.1103/PhysRevD.94.013004} {\bibfield  {journal} {\bibinfo  {journal} {Phys. Rev. D}\ }\textbf {\bibinfo {volume} {94}},\ \bibinfo {pages} {013004} (\bibinfo {year} {2016})}\BibitemShut {NoStop}%
\bibitem [{\citenamefont {King}\ and\ \citenamefont {Elkina}(2016)}]{King_2016_Vacuum}%
  \BibitemOpen
  \bibfield  {author} {\bibinfo {author} {\bibfnamefont {B.}~\bibnamefont {King}}\ and\ \bibinfo {author} {\bibfnamefont {N.}~\bibnamefont {Elkina}},\ }\bibfield  {title} {\bibinfo {title} {Vacuum birefringence in high-energy laser-electron collisions},\ }\href {https://doi.org/10.1103/physreva.94.062102} {\bibfield  {journal} {\bibinfo  {journal} {Phys. Rev. A}\ }\textbf {\bibinfo {volume} {94}},\ \bibinfo {pages} {062102} (\bibinfo {year} {2016})}\BibitemShut {NoStop}%
\bibitem [{\citenamefont {Shen}\ \emph {et~al.}(2018)\citenamefont {Shen}, \citenamefont {Bu}, \citenamefont {Xu}, \citenamefont {Xu}, \citenamefont {Ji}, \citenamefont {Li},\ and\ \citenamefont {Xu}}]{Shen_2018}%
  \BibitemOpen
  \bibfield  {author} {\bibinfo {author} {\bibfnamefont {B.}~\bibnamefont {Shen}}, \bibinfo {author} {\bibfnamefont {Z.}~\bibnamefont {Bu}}, \bibinfo {author} {\bibfnamefont {J.}~\bibnamefont {Xu}}, \bibinfo {author} {\bibfnamefont {T.}~\bibnamefont {Xu}}, \bibinfo {author} {\bibfnamefont {L.}~\bibnamefont {Ji}}, \bibinfo {author} {\bibfnamefont {R.}~\bibnamefont {Li}},\ and\ \bibinfo {author} {\bibfnamefont {Z.}~\bibnamefont {Xu}},\ }\bibfield  {title} {\bibinfo {title} {Exploring vacuum birefringence based on a 100 {PW} laser and an x-ray free electron laser beam},\ }\href {https://doi.org/10.1088/1361-6587/aaa7fb} {\bibfield  {journal} {\bibinfo  {journal} {Plasma Phys. Control. Fusion}\ }\textbf {\bibinfo {volume} {60}},\ \bibinfo {pages} {044002} (\bibinfo {year} {2018})}\BibitemShut {NoStop}%
\bibitem [{\citenamefont {Seino}\ \emph {et~al.}(2020)\citenamefont {Seino}, \citenamefont {Inada}, \citenamefont {Yamazaki}, \citenamefont {Namba},\ and\ \citenamefont {Asai}}]{Seino2020}%
  \BibitemOpen
  \bibfield  {author} {\bibinfo {author} {\bibfnamefont {Y.}~\bibnamefont {Seino}}, \bibinfo {author} {\bibfnamefont {T.}~\bibnamefont {Inada}}, \bibinfo {author} {\bibfnamefont {T.}~\bibnamefont {Yamazaki}}, \bibinfo {author} {\bibfnamefont {T.}~\bibnamefont {Namba}},\ and\ \bibinfo {author} {\bibfnamefont {S.}~\bibnamefont {Asai}},\ }\bibfield  {title} {\bibinfo {title} {{New estimation of the curvature effect for the X-ray vacuum diffraction induced by an intense laser field}},\ }\href {https://doi.org/10.1093/ptep/ptaa084} {\bibfield  {journal} {\bibinfo  {journal} {Prog. Theor. Exp. Phys.}\ }\textbf {\bibinfo {volume} {2020}},\ \bibinfo {pages} {073C02} (\bibinfo {year} {2020})}\BibitemShut {NoStop}%
\bibitem [{\citenamefont {Bragin}\ \emph {et~al.}(2017)\citenamefont {Bragin}, \citenamefont {Meuren}, \citenamefont {Keitel},\ and\ \citenamefont {Piazza}}]{Bragin_2017}%
  \BibitemOpen
  \bibfield  {author} {\bibinfo {author} {\bibfnamefont {S.}~\bibnamefont {Bragin}}, \bibinfo {author} {\bibfnamefont {S.}~\bibnamefont {Meuren}}, \bibinfo {author} {\bibfnamefont {C.~H.}\ \bibnamefont {Keitel}},\ and\ \bibinfo {author} {\bibfnamefont {A.~D.}\ \bibnamefont {Piazza}},\ }\bibfield  {title} {\bibinfo {title} {High-{E}nergy {V}acuum {B}irefringence and {D}ichroism in an {U}ltrastrong {L}aser {F}ield},\ }\href {https://doi.org/10.1103/physrevlett.119.250403} {\bibfield  {journal} {\bibinfo  {journal} {Phys. Rev. Lett.}\ }\textbf {\bibinfo {volume} {119}},\ \bibinfo {pages} {250403} (\bibinfo {year} {2017})}\BibitemShut {NoStop}%
\bibitem [{\citenamefont {Nakamiya}\ and\ \citenamefont {Homma}(2017)}]{Nakamiya_2017}%
  \BibitemOpen
  \bibfield  {author} {\bibinfo {author} {\bibfnamefont {Y.}~\bibnamefont {Nakamiya}}\ and\ \bibinfo {author} {\bibfnamefont {K.}~\bibnamefont {Homma}},\ }\bibfield  {title} {\bibinfo {title} {Probing vacuum birefringence under a high-intensity laser field with gamma-ray polarimetry at the {GeV} scale},\ }\href {https://doi.org/10.1103/physrevd.96.053002} {\bibfield  {journal} {\bibinfo  {journal} {Phys. Rev. D}\ }\textbf {\bibinfo {volume} {96}},\ \bibinfo {pages} {053002} (\bibinfo {year} {2017})}\BibitemShut {NoStop}%
\bibitem [{\citenamefont {Karbstein}(2021)}]{Karbstein2021}%
  \BibitemOpen
  \bibfield  {author} {\bibinfo {author} {\bibfnamefont {F.}~\bibnamefont {Karbstein}},\ }\bibfield  {title} {\bibinfo {title} {{Vacuum} {B}irefringence at the {G}amma {F}actory},\ }\href {https://doi.org/https://doi.org/10.1002/andp.202100137} {\bibfield  {journal} {\bibinfo  {journal} {Ann. Phys.}\ }\textbf {\bibinfo {volume} {2021}},\ \bibinfo {pages} {2100137} (\bibinfo {year} {2021})}\BibitemShut {NoStop}%
\bibitem [{\citenamefont {Berestetskii}\ \emph {et~al.}(1982)\citenamefont {Berestetskii}, , \citenamefont {Lifshitz},\ and\ \citenamefont {Pitevskii}}]{Landau_4}%
  \BibitemOpen
  \bibfield  {author} {\bibinfo {author} {\bibfnamefont {V.~B.}\ \bibnamefont {Berestetskii}}, , \bibinfo {author} {\bibfnamefont {E.~M.}\ \bibnamefont {Lifshitz}},\ and\ \bibinfo {author} {\bibfnamefont {L.~P.}\ \bibnamefont {Pitevskii}},\ }\href@noop {} {\emph {\bibinfo {title} {Quantum electrodynamics}}}\ (\bibinfo  {publisher} {Pergamon, Oxford},\ \bibinfo {year} {1982})\BibitemShut {NoStop}%
\bibitem [{\citenamefont {McMaster}(1961)}]{McMaster1961}%
  \BibitemOpen
  \bibfield  {author} {\bibinfo {author} {\bibfnamefont {W.~H.}\ \bibnamefont {McMaster}},\ }\bibfield  {title} {\bibinfo {title} {{Matrix Representation of Polarization}},\ }\href {https://doi.org/10.1103/revmodphys.33.8} {\bibfield  {journal} {\bibinfo  {journal} {Rev. Mod. Phys.}\ }\textbf {\bibinfo {volume} {33}},\ \bibinfo {pages} {8} (\bibinfo {year} {1961})}\BibitemShut {NoStop}%
\bibitem [{sup()}]{supplemental}%
  \BibitemOpen
  \href@noop {} {}\bibinfo {note} {See supplemental materials for details. It includes the derivation of the refractive index $n(\omega_\gamma)$, simulation results of ideal probe photon beam, different wakefield driver, laser intensity $a_0$ and estimations of $\gamma$-photon scatterings, etc..}\BibitemShut {Stop}%
\bibitem [{\citenamefont {Schubert}(2000)}]{Schubert_2000_vacuum}%
  \BibitemOpen
  \bibfield  {author} {\bibinfo {author} {\bibfnamefont {C.}~\bibnamefont {Schubert}},\ }\bibfield  {title} {\bibinfo {title} {{Vacuum polarization tensors in constant electromagnetic fields. Part 1.}},\ }\href {https://doi.org/10.1016/S0550-3213(00)00423-5} {\bibfield  {journal} {\bibinfo  {journal} {Nucl. Phys. B}\ }\textbf {\bibinfo {volume} {585}},\ \bibinfo {pages} {407} (\bibinfo {year} {2000})}\BibitemShut {NoStop}%
\bibitem [{\citenamefont {Shore}(2007)}]{Shore_2007_Superluminality}%
  \BibitemOpen
  \bibfield  {author} {\bibinfo {author} {\bibfnamefont {G.~M.}\ \bibnamefont {Shore}},\ }\bibfield  {title} {\bibinfo {title} {{Superluminality and {UV} completion}},\ }\href {https://doi.org/10.1016/j.nuclphysb.2007.03.034} {\bibfield  {journal} {\bibinfo  {journal} {Nucl. Phys. B}\ }\textbf {\bibinfo {volume} {778}},\ \bibinfo {pages} {219} (\bibinfo {year} {2007})}\BibitemShut {NoStop}%
\bibitem [{wol()}]{wolfram}%
  \BibitemOpen
  \href {http://mathworld.wolfram.com/AiryFunctions.html} {\bibinfo {title} {{Wolfram} {MathWorld}: {Airy} {Function}}},\ \bibinfo {howpublished} {http://mathworld.wolfram.com/AiryFunctions.html}\BibitemShut {NoStop}%
\bibitem [{\citenamefont {Baier}\ and\ \citenamefont {Breitenlohner}(1967)}]{Baier_1967_vacuum}%
  \BibitemOpen
  \bibfield  {author} {\bibinfo {author} {\bibfnamefont {R.}~\bibnamefont {Baier}}\ and\ \bibinfo {author} {\bibfnamefont {P.}~\bibnamefont {Breitenlohner}},\ }\bibfield  {title} {\bibinfo {title} {{The {V}acuum {R}efraction {I}ndex in the {P}resence of {E}xternal {F}ields}},\ }\href {https://doi.org/10.1007/BF02712312} {\bibfield  {journal} {\bibinfo  {journal} {Nuovo Cim. B}\ }\textbf {\bibinfo {volume} {47}},\ \bibinfo {pages} {117} (\bibinfo {year} {1967})}\BibitemShut {NoStop}%
\bibitem [{\citenamefont {Dinu}\ \emph {et~al.}(2014)\citenamefont {Dinu}, \citenamefont {Heinzl}, \citenamefont {Ilderton}, \citenamefont {Marklund},\ and\ \citenamefont {Torgrimsson}}]{Dinu2014}%
  \BibitemOpen
  \bibfield  {author} {\bibinfo {author} {\bibfnamefont {V.}~\bibnamefont {Dinu}}, \bibinfo {author} {\bibfnamefont {T.}~\bibnamefont {Heinzl}}, \bibinfo {author} {\bibfnamefont {A.}~\bibnamefont {Ilderton}}, \bibinfo {author} {\bibfnamefont {M.}~\bibnamefont {Marklund}},\ and\ \bibinfo {author} {\bibfnamefont {G.}~\bibnamefont {Torgrimsson}},\ }\bibfield  {title} {\bibinfo {title} {Vacuum refractive indices and helicity flip in strong-field {QED}},\ }\href {https://doi.org/10.1103/physrevd.89.125003} {\bibfield  {journal} {\bibinfo  {journal} {Phys. Rev. D}\ }\textbf {\bibinfo {volume} {89}},\ \bibinfo {pages} {125003} (\bibinfo {year} {2014})}\BibitemShut {NoStop}%
\bibitem [{\citenamefont {Arber}\ \emph {et~al.}(2015)\citenamefont {Arber}, \citenamefont {Bennett}, \citenamefont {Brady}, \citenamefont {Lawrence-Douglas}, \citenamefont {Ramsay}, \citenamefont {Sircombe}, \citenamefont {Gillies}, \citenamefont {Evans}, \citenamefont {Schmitz}, \citenamefont {Bell},\ and\ \citenamefont {Ridgers}}]{Arber2015}%
  \BibitemOpen
  \bibfield  {author} {\bibinfo {author} {\bibfnamefont {T.~D.}\ \bibnamefont {Arber}}, \bibinfo {author} {\bibfnamefont {K.}~\bibnamefont {Bennett}}, \bibinfo {author} {\bibfnamefont {C.~S.}\ \bibnamefont {Brady}}, \bibinfo {author} {\bibfnamefont {A.}~\bibnamefont {Lawrence-Douglas}}, \bibinfo {author} {\bibfnamefont {M.~G.}\ \bibnamefont {Ramsay}}, \bibinfo {author} {\bibfnamefont {N.~J.}\ \bibnamefont {Sircombe}}, \bibinfo {author} {\bibfnamefont {P.}~\bibnamefont {Gillies}}, \bibinfo {author} {\bibfnamefont {R.~G.}\ \bibnamefont {Evans}}, \bibinfo {author} {\bibfnamefont {H.}~\bibnamefont {Schmitz}}, \bibinfo {author} {\bibfnamefont {A.~R.}\ \bibnamefont {Bell}},\ and\ \bibinfo {author} {\bibfnamefont {C.~P.}\ \bibnamefont {Ridgers}},\ }\bibfield  {title} {\bibinfo {title} {Contemporary particle-in-cell approach to laser-plasma modelling},\ }\href {https://doi.org/10.1088/0741-3335/57/11/113001} {\bibfield  {journal} {\bibinfo  {journal} {Plasma Phys. Control. Fusion}\ }\textbf {\bibinfo {volume} {57}},\ \bibinfo {pages} {113001} (\bibinfo {year} {2015})}\BibitemShut {NoStop}%
\bibitem [{\citenamefont {Xue}\ \emph {et~al.}(2020)\citenamefont {Xue}, \citenamefont {Dou}, \citenamefont {Wan}, \citenamefont {Yu}, \citenamefont {Wang}, \citenamefont {Ren}, \citenamefont {Zhao}, \citenamefont {Zhao}, \citenamefont {Xu},\ and\ \citenamefont {Li}}]{Xue2020}%
  \BibitemOpen
  \bibfield  {author} {\bibinfo {author} {\bibfnamefont {K.}~\bibnamefont {Xue}}, \bibinfo {author} {\bibfnamefont {Z.-K.}\ \bibnamefont {Dou}}, \bibinfo {author} {\bibfnamefont {F.}~\bibnamefont {Wan}}, \bibinfo {author} {\bibfnamefont {T.-P.}\ \bibnamefont {Yu}}, \bibinfo {author} {\bibfnamefont {W.-M.}\ \bibnamefont {Wang}}, \bibinfo {author} {\bibfnamefont {J.-R.}\ \bibnamefont {Ren}}, \bibinfo {author} {\bibfnamefont {Q.}~\bibnamefont {Zhao}}, \bibinfo {author} {\bibfnamefont {Y.-T.}\ \bibnamefont {Zhao}}, \bibinfo {author} {\bibfnamefont {Z.-F.}\ \bibnamefont {Xu}},\ and\ \bibinfo {author} {\bibfnamefont {J.-X.}\ \bibnamefont {Li}},\ }\bibfield  {title} {\bibinfo {title} {Generation of highly-polarized high-energy brilliant $\gamma$-rays via laser-plasma interaction},\ }\href {https://doi.org/10.1063/5.0007734} {\bibfield  {journal} {\bibinfo  {journal} {Matter Radiat. Extremes}\ }\textbf {\bibinfo {volume} {5}},\ \bibinfo {pages} {054402} (\bibinfo {year} {2020})}\BibitemShut {NoStop}%
\bibitem [{\citenamefont {Wang}\ \emph {et~al.}(2021)\citenamefont {Wang}, \citenamefont {Feng}, \citenamefont {Ke}, \citenamefont {Yu}, \citenamefont {Xu}, \citenamefont {Qi}, \citenamefont {Chen}, \citenamefont {Qin}, \citenamefont {Zhang}, \citenamefont {Fang}, \citenamefont {Liu}, \citenamefont {Jiang}, \citenamefont {Wang}, \citenamefont {Wang}, \citenamefont {Yang}, \citenamefont {Wu}, \citenamefont {Leng}, \citenamefont {Liu}, \citenamefont {Li},\ and\ \citenamefont {Xu}}]{Wang_2021_Free}%
  \BibitemOpen
  \bibfield  {author} {\bibinfo {author} {\bibfnamefont {W.}~\bibnamefont {Wang}}, \bibinfo {author} {\bibfnamefont {K.}~\bibnamefont {Feng}}, \bibinfo {author} {\bibfnamefont {L.}~\bibnamefont {Ke}}, \bibinfo {author} {\bibfnamefont {C.}~\bibnamefont {Yu}}, \bibinfo {author} {\bibfnamefont {Y.}~\bibnamefont {Xu}}, \bibinfo {author} {\bibfnamefont {R.}~\bibnamefont {Qi}}, \bibinfo {author} {\bibfnamefont {Y.}~\bibnamefont {Chen}}, \bibinfo {author} {\bibfnamefont {Z.}~\bibnamefont {Qin}}, \bibinfo {author} {\bibfnamefont {Z.}~\bibnamefont {Zhang}}, \bibinfo {author} {\bibfnamefont {M.}~\bibnamefont {Fang}}, \bibinfo {author} {\bibfnamefont {J.}~\bibnamefont {Liu}}, \bibinfo {author} {\bibfnamefont {K.}~\bibnamefont {Jiang}}, \bibinfo {author} {\bibfnamefont {H.}~\bibnamefont {Wang}}, \bibinfo {author} {\bibfnamefont {C.}~\bibnamefont {Wang}}, \bibinfo {author} {\bibfnamefont {X.}~\bibnamefont {Yang}}, \bibinfo {author} {\bibfnamefont {F.}~\bibnamefont {Wu}}, \bibinfo {author} {\bibfnamefont {Y.}~\bibnamefont {Leng}}, \bibinfo {author} {\bibfnamefont {J.}~\bibnamefont {Liu}}, \bibinfo {author} {\bibfnamefont {R.}~\bibnamefont {Li}},\ and\ \bibinfo {author} {\bibfnamefont {Z.}~\bibnamefont {Xu}},\ }\bibfield  {title} {\bibinfo {title} {Free-electron lasing at 27 nanometres based on a laser wakefield accelerator},\ }\href {https://doi.org/10.1038/s41586-021-03678-x} {\bibfield  {journal} {\bibinfo  {journal} {Nature}\ }\textbf {\bibinfo {volume} {595}},\ \bibinfo {pages} {516} (\bibinfo {year} {2021})}\BibitemShut {NoStop}%
\bibitem [{\citenamefont {Gonsalves}\ \emph {et~al.}(2019)\citenamefont {Gonsalves}, \citenamefont {Nakamura}, \citenamefont {Daniels}, \citenamefont {Benedetti}, \citenamefont {Pieronek}, \citenamefont {de~Raadt}, \citenamefont {Steinke}, \citenamefont {Bin}, \citenamefont {Bulanov}, \citenamefont {van Tilborg}, \citenamefont {Geddes}, \citenamefont {Schroeder}, \citenamefont {T\'oth}, \citenamefont {Esarey}, \citenamefont {Swanson}, \citenamefont {Fan-Chiang}, \citenamefont {Bagdasarov}, \citenamefont {Bobrova}, \citenamefont {Gasilov}, \citenamefont {Korn}, \citenamefont {Sasorov},\ and\ \citenamefont {Leemans}}]{Gonsalves2019}%
  \BibitemOpen
  \bibfield  {author} {\bibinfo {author} {\bibfnamefont {A.~J.}\ \bibnamefont {Gonsalves}}, \bibinfo {author} {\bibfnamefont {K.}~\bibnamefont {Nakamura}}, \bibinfo {author} {\bibfnamefont {J.}~\bibnamefont {Daniels}}, \bibinfo {author} {\bibfnamefont {C.}~\bibnamefont {Benedetti}}, \bibinfo {author} {\bibfnamefont {C.}~\bibnamefont {Pieronek}}, \bibinfo {author} {\bibfnamefont {T.~C.~H.}\ \bibnamefont {de~Raadt}}, \bibinfo {author} {\bibfnamefont {S.}~\bibnamefont {Steinke}}, \bibinfo {author} {\bibfnamefont {J.~H.}\ \bibnamefont {Bin}}, \bibinfo {author} {\bibfnamefont {S.~S.}\ \bibnamefont {Bulanov}}, \bibinfo {author} {\bibfnamefont {J.}~\bibnamefont {van Tilborg}}, \bibinfo {author} {\bibfnamefont {C.~G.~R.}\ \bibnamefont {Geddes}}, \bibinfo {author} {\bibfnamefont {C.~B.}\ \bibnamefont {Schroeder}}, \bibinfo {author} {\bibfnamefont {C.}~\bibnamefont {T\'oth}}, \bibinfo {author} {\bibfnamefont {E.}~\bibnamefont {Esarey}}, \bibinfo {author} {\bibfnamefont {K.}~\bibnamefont {Swanson}}, \bibinfo {author} {\bibfnamefont {L.}~\bibnamefont {Fan-Chiang}}, \bibinfo {author} {\bibfnamefont {G.}~\bibnamefont {Bagdasarov}}, \bibinfo {author} {\bibfnamefont {N.}~\bibnamefont {Bobrova}}, \bibinfo {author} {\bibfnamefont {V.}~\bibnamefont {Gasilov}}, \bibinfo {author} {\bibfnamefont {G.}~\bibnamefont {Korn}}, \bibinfo {author} {\bibfnamefont {P.}~\bibnamefont {Sasorov}},\ and\ \bibinfo {author} {\bibfnamefont {W.~P.}\ \bibnamefont {Leemans}},\ }\bibfield  {title} {\bibinfo {title} {{P}etawatt {L}aser {G}uiding and {E}lectron {B}eam {A}cceleration to 8 {G}ev in a {L}aser-{H}eated {C}apillary {D}ischarge {W}aveguide},\ }\href {https://doi.org/10.1103/PhysRevLett.122.084801} {\bibfield  {journal} {\bibinfo  {journal} {Phys. Rev. Lett.}\ }\textbf {\bibinfo {volume} {122}},\ \bibinfo {pages} {084801} (\bibinfo {year} {2019})}\BibitemShut {NoStop}%
\bibitem [{\citenamefont {Ur}(2015)}]{Ur_2015}%
  \BibitemOpen
  \bibfield  {author} {\bibinfo {author} {\bibfnamefont {C.~A.}\ \bibnamefont {Ur}},\ }\bibfield  {title} {\bibinfo {title} {Gamma beam system at {ELI-NP}},\ }\href {https://doi.org/10.1063/1.4909580} {\bibfield  {journal} {\bibinfo  {journal} {AIP Conference Proceedings}\ }\textbf {\bibinfo {volume} {1645}},\ \bibinfo {pages} {237} (\bibinfo {year} {2015})}\BibitemShut {NoStop}%
\bibitem [{\citenamefont {Li}\ \emph {et~al.}(2020)\citenamefont {Li}, \citenamefont {Shaisultanov}, \citenamefont {Chen}, \citenamefont {Wan}, \citenamefont {Hatsagortsyan}, \citenamefont {Keitel},\ and\ \citenamefont {Li}}]{Li_2020_Polarized}%
  \BibitemOpen
  \bibfield  {author} {\bibinfo {author} {\bibfnamefont {Y.-F.}\ \bibnamefont {Li}}, \bibinfo {author} {\bibfnamefont {R.}~\bibnamefont {Shaisultanov}}, \bibinfo {author} {\bibfnamefont {Y.-Y.}\ \bibnamefont {Chen}}, \bibinfo {author} {\bibfnamefont {F.}~\bibnamefont {Wan}}, \bibinfo {author} {\bibfnamefont {K.~Z.}\ \bibnamefont {Hatsagortsyan}}, \bibinfo {author} {\bibfnamefont {C.~H.}\ \bibnamefont {Keitel}},\ and\ \bibinfo {author} {\bibfnamefont {J.-X.}\ \bibnamefont {Li}},\ }\bibfield  {title} {\bibinfo {title} {Polarized {U}ltrashort {B}rilliant {M}ulti-{GeV} $\gamma$-{R}ays via {S}ingle-{S}hot {L}aser-{E}lectron {I}nteraction},\ }\href {https://doi.org/10.1103/physrevlett.124.014801} {\bibfield  {journal} {\bibinfo  {journal} {Phys. Rev. Lett.}\ }\textbf {\bibinfo {volume} {124}},\ \bibinfo {pages} {014801} (\bibinfo {year} {2020})}\BibitemShut {NoStop}%
\bibitem [{\citenamefont {Sampath}\ \emph {et~al.}(2021)\citenamefont {Sampath}, \citenamefont {Davoine}, \citenamefont {Corde}, \citenamefont {Gremillet}, \citenamefont {Gilljohann}, \citenamefont {Sangal}, \citenamefont {Keitel}, \citenamefont {Ariniello}, \citenamefont {Cary}, \citenamefont {Ekerfelt}, \citenamefont {Emma}, \citenamefont {Fiuza}, \citenamefont {Fujii}, \citenamefont {Hogan}, \citenamefont {Joshi}, \citenamefont {Knetsch}, \citenamefont {Kononenko}, \citenamefont {Lee}, \citenamefont {Litos}, \citenamefont {Marsh}, \citenamefont {Nie}, \citenamefont {O'Shea}, \citenamefont {Peterson}, \citenamefont {Claveria}, \citenamefont {Storey}, \citenamefont {Wu}, \citenamefont {Xu}, \citenamefont {Zhang},\ and\ \citenamefont {Tamburini}}]{Sampath_2021}%
  \BibitemOpen
  \bibfield  {author} {\bibinfo {author} {\bibfnamefont {A.}~\bibnamefont {Sampath}}, \bibinfo {author} {\bibfnamefont {X.}~\bibnamefont {Davoine}}, \bibinfo {author} {\bibfnamefont {S.}~\bibnamefont {Corde}}, \bibinfo {author} {\bibfnamefont {L.}~\bibnamefont {Gremillet}}, \bibinfo {author} {\bibfnamefont {M.}~\bibnamefont {Gilljohann}}, \bibinfo {author} {\bibfnamefont {M.}~\bibnamefont {Sangal}}, \bibinfo {author} {\bibfnamefont {C.~H.}\ \bibnamefont {Keitel}}, \bibinfo {author} {\bibfnamefont {R.}~\bibnamefont {Ariniello}}, \bibinfo {author} {\bibfnamefont {J.}~\bibnamefont {Cary}}, \bibinfo {author} {\bibfnamefont {H.}~\bibnamefont {Ekerfelt}}, \bibinfo {author} {\bibfnamefont {C.}~\bibnamefont {Emma}}, \bibinfo {author} {\bibfnamefont {F.}~\bibnamefont {Fiuza}}, \bibinfo {author} {\bibfnamefont {H.}~\bibnamefont {Fujii}}, \bibinfo {author} {\bibfnamefont {M.}~\bibnamefont {Hogan}}, \bibinfo {author} {\bibfnamefont {C.}~\bibnamefont {Joshi}}, \bibinfo {author} {\bibfnamefont {A.}~\bibnamefont {Knetsch}}, \bibinfo {author} {\bibfnamefont {O.}~\bibnamefont {Kononenko}}, \bibinfo {author} {\bibfnamefont {V.}~\bibnamefont {Lee}}, \bibinfo {author} {\bibfnamefont {M.}~\bibnamefont {Litos}}, \bibinfo {author} {\bibfnamefont {K.}~\bibnamefont {Marsh}}, \bibinfo {author} {\bibfnamefont {Z.}~\bibnamefont {Nie}}, \bibinfo {author} {\bibfnamefont {B.}~\bibnamefont {O'Shea}}, \bibinfo {author} {\bibfnamefont {J.~R.}\ \bibnamefont {Peterson}}, \bibinfo {author} {\bibfnamefont {P.~S.~M.}\ \bibnamefont {Claveria}}, \bibinfo {author} {\bibfnamefont {D.}~\bibnamefont {Storey}}, \bibinfo {author} {\bibfnamefont {Y.}~\bibnamefont {Wu}}, \bibinfo {author} {\bibfnamefont {X.}~\bibnamefont {Xu}}, \bibinfo {author} {\bibfnamefont {C.}~\bibnamefont {Zhang}},\ and\ \bibinfo {author} {\bibfnamefont {M.}~\bibnamefont {Tamburini}},\ }\bibfield  {title} {\bibinfo {title} {Extremely dense gamma-ray pulses in electron beam-multifoil collisions},\ }\href {https://doi.org/10.1103/PhysRevLett.126.064801} {\bibfield  {journal} {\bibinfo  {journal} {Phys. Rev. Lett.}\ }\textbf {\bibinfo {volume} {126}},\ \bibinfo {pages} {064801} (\bibinfo {year} {2021})}\BibitemShut {NoStop}%
\bibitem [{\citenamefont {Esarey}\ \emph {et~al.}(2009)\citenamefont {Esarey}, \citenamefont {Schroeder},\ and\ \citenamefont {Leemans}}]{Esarey2009}%
  \BibitemOpen
  \bibfield  {author} {\bibinfo {author} {\bibfnamefont {E.}~\bibnamefont {Esarey}}, \bibinfo {author} {\bibfnamefont {C.~B.}\ \bibnamefont {Schroeder}},\ and\ \bibinfo {author} {\bibfnamefont {W.~P.}\ \bibnamefont {Leemans}},\ }\bibfield  {title} {\bibinfo {title} {Physics of laser-driven plasma-based electron accelerators},\ }\href {https://doi.org/10.1103/revmodphys.81.1229} {\bibfield  {journal} {\bibinfo  {journal} {Rev. Mod. Phys.}\ }\textbf {\bibinfo {volume} {81}},\ \bibinfo {pages} {1229} (\bibinfo {year} {2009})}\BibitemShut {NoStop}%
\bibitem [{\citenamefont {Schopper}(1958)}]{Schopper_1958}%
  \BibitemOpen
  \bibfield  {author} {\bibinfo {author} {\bibfnamefont {H.}~\bibnamefont {Schopper}},\ }\bibfield  {title} {\bibinfo {title} {Measurement of circular polarization of $\gamma$-rays},\ }\href {https://doi.org/10.1016/0369-643X(58)90018-5} {\bibfield  {journal} {\bibinfo  {journal} {Nucl. Instrum.}\ }\textbf {\bibinfo {volume} {3}},\ \bibinfo {pages} {158} (\bibinfo {year} {1958})}\BibitemShut {NoStop}%
\bibitem [{\citenamefont {Fukuda}\ \emph {et~al.}(2003)\citenamefont {Fukuda}, \citenamefont {Aoki}, \citenamefont {Dobashi}, \citenamefont {Hirose}, \citenamefont {Iimura}, \citenamefont {Kurihara}, \citenamefont {Okugi}, \citenamefont {Omori}, \citenamefont {Sakai}, \citenamefont {Urakawa},\ and\ \citenamefont {Washio}}]{Fukuda_2003}%
  \BibitemOpen
  \bibfield  {author} {\bibinfo {author} {\bibfnamefont {M.}~\bibnamefont {Fukuda}}, \bibinfo {author} {\bibfnamefont {T.}~\bibnamefont {Aoki}}, \bibinfo {author} {\bibfnamefont {K.}~\bibnamefont {Dobashi}}, \bibinfo {author} {\bibfnamefont {T.}~\bibnamefont {Hirose}}, \bibinfo {author} {\bibfnamefont {T.}~\bibnamefont {Iimura}}, \bibinfo {author} {\bibfnamefont {Y.}~\bibnamefont {Kurihara}}, \bibinfo {author} {\bibfnamefont {T.}~\bibnamefont {Okugi}}, \bibinfo {author} {\bibfnamefont {T.}~\bibnamefont {Omori}}, \bibinfo {author} {\bibfnamefont {I.}~\bibnamefont {Sakai}}, \bibinfo {author} {\bibfnamefont {J.}~\bibnamefont {Urakawa}},\ and\ \bibinfo {author} {\bibfnamefont {M.}~\bibnamefont {Washio}},\ }\bibfield  {title} {\bibinfo {title} {Polarimetry of short-pulse gamma rays produced through inverse compton scattering of circularly polarized laser beams},\ }\href {https://doi.org/10.1103/physrevlett.91.164801} {\bibfield  {journal} {\bibinfo  {journal} {Phys. Rev. Lett.}\ }\textbf {\bibinfo {volume} {91}},\ \bibinfo {pages} {164801} (\bibinfo {year} {2003})}\BibitemShut {NoStop}%
\bibitem [{\citenamefont {Tashenov}(2011)}]{Tashenov_2011}%
  \BibitemOpen
  \bibfield  {author} {\bibinfo {author} {\bibfnamefont {S.}~\bibnamefont {Tashenov}},\ }\bibfield  {title} {\bibinfo {title} {Circular polarimetry with gamma-ray tracking detectors},\ }\href {https://doi.org/10.1016/j.nima.2011.03.011} {\bibfield  {journal} {\bibinfo  {journal} {Nucl. Instrum. Meth. A.}\ }\textbf {\bibinfo {volume} {640}},\ \bibinfo {pages} {164} (\bibinfo {year} {2011})}\BibitemShut {NoStop}%
\bibitem [{\citenamefont {Ilie}(2019)}]{Ilie_2019}%
  \BibitemOpen
  \bibfield  {author} {\bibinfo {author} {\bibfnamefont {C.}~\bibnamefont {Ilie}},\ }\bibfield  {title} {\bibinfo {title} {Gamma-ray polarimetry: A new window for the nonthermal universe},\ }\href {https://doi.org/10.1088/1538-3873/ab2a3a} {\bibfield  {journal} {\bibinfo  {journal} {Publications of the Astronomical Society of the Pacific}\ }\textbf {\bibinfo {volume} {131}},\ \bibinfo {pages} {111001} (\bibinfo {year} {2019})}\BibitemShut {NoStop}%
\bibitem [{\citenamefont {Kostyukov}\ \emph {et~al.}(2004)\citenamefont {Kostyukov}, \citenamefont {Pukhov},\ and\ \citenamefont {Kiselev}}]{Kostyukov2004}%
  \BibitemOpen
  \bibfield  {author} {\bibinfo {author} {\bibfnamefont {I.}~\bibnamefont {Kostyukov}}, \bibinfo {author} {\bibfnamefont {A.}~\bibnamefont {Pukhov}},\ and\ \bibinfo {author} {\bibfnamefont {S.}~\bibnamefont {Kiselev}},\ }\bibfield  {title} {\bibinfo {title} {Phenomenological theory of laser-plasma interaction in {\textquotedblleft}bubble{\textquotedblright} regime},\ }\href {https://doi.org/10.1063/1.1799371} {\bibfield  {journal} {\bibinfo  {journal} {Phys. Plasmas}\ }\textbf {\bibinfo {volume} {11}},\ \bibinfo {pages} {5256} (\bibinfo {year} {2004})}\BibitemShut {NoStop}%
\bibitem [{\citenamefont {Lu}\ \emph {et~al.}(2006)\citenamefont {Lu}, \citenamefont {Huang}, \citenamefont {Zhou}, \citenamefont {Mori},\ and\ \citenamefont {Katsouleas}}]{Lu2006}%
  \BibitemOpen
  \bibfield  {author} {\bibinfo {author} {\bibfnamefont {W.}~\bibnamefont {Lu}}, \bibinfo {author} {\bibfnamefont {C.}~\bibnamefont {Huang}}, \bibinfo {author} {\bibfnamefont {M.}~\bibnamefont {Zhou}}, \bibinfo {author} {\bibfnamefont {W.~B.}\ \bibnamefont {Mori}},\ and\ \bibinfo {author} {\bibfnamefont {T.}~\bibnamefont {Katsouleas}},\ }\bibfield  {title} {\bibinfo {title} {{Nonlinear Theory for Relativistic Plasma Wakefields in the Blowout Regime}},\ }\href {https://doi.org/10.1103/physrevlett.96.165002} {\bibfield  {journal} {\bibinfo  {journal} {Phys. Rev. Lett.}\ }\textbf {\bibinfo {volume} {96}},\ \bibinfo {pages} {165002} (\bibinfo {year} {2006})}\BibitemShut {NoStop}%
\bibitem [{\citenamefont {Corde}\ \emph {et~al.}(2013)\citenamefont {Corde}, \citenamefont {Phuoc}, \citenamefont {Lambert}, \citenamefont {Fitour}, \citenamefont {Malka}, \citenamefont {Rousse}, \citenamefont {Beck},\ and\ \citenamefont {Lefebvre}}]{Corde2013}%
  \BibitemOpen
  \bibfield  {author} {\bibinfo {author} {\bibfnamefont {S.}~\bibnamefont {Corde}}, \bibinfo {author} {\bibfnamefont {K.~T.}\ \bibnamefont {Phuoc}}, \bibinfo {author} {\bibfnamefont {G.}~\bibnamefont {Lambert}}, \bibinfo {author} {\bibfnamefont {R.}~\bibnamefont {Fitour}}, \bibinfo {author} {\bibfnamefont {V.}~\bibnamefont {Malka}}, \bibinfo {author} {\bibfnamefont {A.}~\bibnamefont {Rousse}}, \bibinfo {author} {\bibfnamefont {A.}~\bibnamefont {Beck}},\ and\ \bibinfo {author} {\bibfnamefont {E.}~\bibnamefont {Lefebvre}},\ }\bibfield  {title} {\bibinfo {title} {Femtosecond x-rays from laser-plasma accelerators},\ }\href {https://doi.org/10.1103/revmodphys.85.1} {\bibfield  {journal} {\bibinfo  {journal} {Rev. Mod. Phys.}\ }\textbf {\bibinfo {volume} {85}},\ \bibinfo {pages} {1} (\bibinfo {year} {2013})}\BibitemShut {NoStop}%
\bibitem [{\citenamefont {King}\ and\ \citenamefont {Tang}(2020)}]{King2020}%
  \BibitemOpen
  \bibfield  {author} {\bibinfo {author} {\bibfnamefont {B.}~\bibnamefont {King}}\ and\ \bibinfo {author} {\bibfnamefont {S.}~\bibnamefont {Tang}},\ }\bibfield  {title} {\bibinfo {title} {Nonlinear compton scattering of polarized photons in plane-wave backgrounds},\ }\href {https://doi.org/10.1103/physreva.102.022809} {\bibfield  {journal} {\bibinfo  {journal} {Phys. Rev. A}\ }\textbf {\bibinfo {volume} {102}},\ \bibinfo {pages} {022809} (\bibinfo {year} {2020})}\BibitemShut {NoStop}%
\bibitem [{\citenamefont {Tang}\ \emph {et~al.}(2020)\citenamefont {Tang}, \citenamefont {King},\ and\ \citenamefont {Hu}}]{Tang2020}%
  \BibitemOpen
  \bibfield  {author} {\bibinfo {author} {\bibfnamefont {S.}~\bibnamefont {Tang}}, \bibinfo {author} {\bibfnamefont {B.}~\bibnamefont {King}},\ and\ \bibinfo {author} {\bibfnamefont {H.}~\bibnamefont {Hu}},\ }\bibfield  {title} {\bibinfo {title} {Highly polarised gamma photons from electron-laser collisions},\ }\href {https://doi.org/10.1016/j.physletb.2020.135701} {\bibfield  {journal} {\bibinfo  {journal} {Phys. Lett. B}\ }\textbf {\bibinfo {volume} {809}},\ \bibinfo {pages} {135701} (\bibinfo {year} {2020})}\BibitemShut {NoStop}%
\bibitem [{\citenamefont {Ritus}(1985)}]{Ritus_1985_Quantum}%
  \BibitemOpen
  \bibfield  {author} {\bibinfo {author} {\bibfnamefont {V.~I.}\ \bibnamefont {Ritus}},\ }\bibfield  {title} {\bibinfo {title} {Quantum effects of the interaction of elementary particles with an intense electromagnetic field},\ }\href {https://doi.org/10.1007/bf01120220} {\bibfield  {journal} {\bibinfo  {journal} {J. Russ. Laser Res.}\ }\textbf {\bibinfo {volume} {6}},\ \bibinfo {pages} {497} (\bibinfo {year} {1985})}\BibitemShut {NoStop}%
\bibitem [{\citenamefont {Baier}\ \emph {et~al.}(1998)\citenamefont {Baier}, \citenamefont {Katkov},\ and\ \citenamefont {Strakhovenko}}]{Baier_1998_Electromagnetic}%
  \BibitemOpen
  \bibfield  {author} {\bibinfo {author} {\bibfnamefont {V.~N.}\ \bibnamefont {Baier}}, \bibinfo {author} {\bibfnamefont {V.~M.}\ \bibnamefont {Katkov}},\ and\ \bibinfo {author} {\bibfnamefont {V.~M.}\ \bibnamefont {Strakhovenko}},\ }\href {https://doi.org/10.1142/2216} {\emph {\bibinfo {title} {Electromagnetic {P}rocesses at {H}igh {E}nergies in {O}riented {S}ingle {C}rystals}}}\ (\bibinfo  {publisher} {{WORLD} {SCIENTIFIC}},\ \bibinfo {year} {1998})\BibitemShut {NoStop}%
\bibitem [{\citenamefont {Bell}\ and\ \citenamefont {Kirk}(2008)}]{Bell_2008_Possibility}%
  \BibitemOpen
  \bibfield  {author} {\bibinfo {author} {\bibfnamefont {A.~R.}\ \bibnamefont {Bell}}\ and\ \bibinfo {author} {\bibfnamefont {J.~G.}\ \bibnamefont {Kirk}},\ }\bibfield  {title} {\bibinfo {title} {Possibility of {P}rolific {P}air {P}roduction with {H}igh-{P}ower {L}asers},\ }\href {https://doi.org/10.1103/physrevlett.101.200403} {\bibfield  {journal} {\bibinfo  {journal} {Phys. Rev. Lett.}\ }\textbf {\bibinfo {volume} {101}},\ \bibinfo {pages} {200403} (\bibinfo {year} {2008})}\BibitemShut {NoStop}%
\bibitem [{\citenamefont {Ridgers}\ \emph {et~al.}(2012)\citenamefont {Ridgers}, \citenamefont {Brady}, \citenamefont {Duclous}, \citenamefont {Kirk}, \citenamefont {Bennett}, \citenamefont {Arber}, \citenamefont {Robinson},\ and\ \citenamefont {Bell}}]{Ridgers_2012_Dense}%
  \BibitemOpen
  \bibfield  {author} {\bibinfo {author} {\bibfnamefont {C.~P.}\ \bibnamefont {Ridgers}}, \bibinfo {author} {\bibfnamefont {C.~S.}\ \bibnamefont {Brady}}, \bibinfo {author} {\bibfnamefont {R.}~\bibnamefont {Duclous}}, \bibinfo {author} {\bibfnamefont {J.~G.}\ \bibnamefont {Kirk}}, \bibinfo {author} {\bibfnamefont {K.}~\bibnamefont {Bennett}}, \bibinfo {author} {\bibfnamefont {T.~D.}\ \bibnamefont {Arber}}, \bibinfo {author} {\bibfnamefont {A.~P.~L.}\ \bibnamefont {Robinson}},\ and\ \bibinfo {author} {\bibfnamefont {A.~R.}\ \bibnamefont {Bell}},\ }\bibfield  {title} {\bibinfo {title} {Dense {E}lectron-{P}ositron {P}lasmas and {U}ltraintense $\gamma$-rays from {L}aser-{I}rradiated {S}olids},\ }\href {https://doi.org/10.1103/physrevlett.108.165006} {\bibfield  {journal} {\bibinfo  {journal} {Phys. Rev. Lett.}\ }\textbf {\bibinfo {volume} {108}},\ \bibinfo {pages} {165006} (\bibinfo {year} {2012})}\BibitemShut {NoStop}%
\bibitem [{nis()}]{nist-xcome}%
  \BibitemOpen
  \href {https://www.physics.nist.gov/PhysRefData/Xcom/html/xcom1.html} {\bibinfo {title} {{NIST XCOM: P}hoton {C}ross {S}ections {D}atabase}},\ \bibinfo {howpublished} {https://www.physics.nist.gov/PhysRefData/Xcom/html/xcom1.html}\BibitemShut {NoStop}%
\bibitem [{\citenamefont {Tsai}(1974)}]{Tsai_1974_Pair}%
  \BibitemOpen
  \bibfield  {author} {\bibinfo {author} {\bibfnamefont {Y.-S.}\ \bibnamefont {Tsai}},\ }\bibfield  {title} {\bibinfo {title} {Pair production and bremsstrahlung of charged leptons},\ }\href {https://doi.org/10.1103/revmodphys.46.815} {\bibfield  {journal} {\bibinfo  {journal} {Rev. Mod. Phys.}\ }\textbf {\bibinfo {volume} {46}},\ \bibinfo {pages} {815} (\bibinfo {year} {1974})}\BibitemShut {NoStop}%
\bibitem [{\citenamefont {Hubbell}(2006)}]{Hubbell_2006_Review}%
  \BibitemOpen
  \bibfield  {author} {\bibinfo {author} {\bibfnamefont {J.~H.}\ \bibnamefont {Hubbell}},\ }\bibfield  {title} {\bibinfo {title} {Review and history of photon cross section calculations},\ }\href {https://doi.org/10.1088/0031-9155/51/13/r15} {\bibfield  {journal} {\bibinfo  {journal} {Phys. Med. Biol.}\ }\textbf {\bibinfo {volume} {51}},\ \bibinfo {pages} {R245} (\bibinfo {year} {2006})}\BibitemShut {NoStop}%
\bibitem [{\citenamefont {Salvat}\ and\ \citenamefont {Fern{\'{a}}ndez-Varea}(2009)}]{Salvat_2009_Overview}%
  \BibitemOpen
  \bibfield  {author} {\bibinfo {author} {\bibfnamefont {F.}~\bibnamefont {Salvat}}\ and\ \bibinfo {author} {\bibfnamefont {J.~M.}\ \bibnamefont {Fern{\'{a}}ndez-Varea}},\ }\bibfield  {title} {\bibinfo {title} {Overview of physical interaction models for photon and electron transport used in {M}onte {C}arlo codes},\ }\href {https://doi.org/10.1088/0026-1394/46/2/s08} {\bibfield  {journal} {\bibinfo  {journal} {Metrologia}\ }\textbf {\bibinfo {volume} {46}},\ \bibinfo {pages} {S112} (\bibinfo {year} {2009})}\BibitemShut {NoStop}%
\bibitem [{\citenamefont {Allison}\ \emph {et~al.}(2016)\citenamefont {Allison}, \citenamefont {Amako}, \citenamefont {Apostolakis}, \citenamefont {Arce}, \citenamefont {Asai}, \citenamefont {Aso}, \citenamefont {Bagli}, \citenamefont {Bagulya}, \citenamefont {Banerjee}, \citenamefont {Barrand}, \citenamefont {Beck}, \citenamefont {Bogdanov}, \citenamefont {Brandt}, \citenamefont {Brown}, \citenamefont {Burkhardt}, \citenamefont {Canal}, \citenamefont {Cano-Ott}, \citenamefont {Chauvie}, \citenamefont {Cho}, \citenamefont {Cirrone}, \citenamefont {Cooperman}, \citenamefont {Cortés-Giraldo}, \citenamefont {Cosmo}, \citenamefont {Cuttone}, \citenamefont {Depaola}, \citenamefont {Desorgher}, \citenamefont {Dong}, \citenamefont {Dotti}, \citenamefont {Elvira}, \citenamefont {Folger}, \citenamefont {Francis}, \citenamefont {Galoyan}, \citenamefont {Garnier}, \citenamefont {Gayer}, \citenamefont {Genser}, \citenamefont {Grichine}, \citenamefont {Guatelli}, \citenamefont {Guèye}, \citenamefont {Gumplinger}, \citenamefont {Howard}, \citenamefont {Hřivnáčová}, \citenamefont {Hwang}, \citenamefont {Incerti}, \citenamefont {Ivanchenko}, \citenamefont {Ivanchenko}, \citenamefont {Jones}, \citenamefont {Jun}, \citenamefont {Kaitaniemi}, \citenamefont {Karakatsanis}, \citenamefont {Karamitros}, \citenamefont {Kelsey}, \citenamefont {Kimura}, \citenamefont {Koi}, \citenamefont {Kurashige}, \citenamefont {Lechner}, \citenamefont {Lee}, \citenamefont {Longo}, \citenamefont {Maire}, \citenamefont {Mancusi}, \citenamefont {Mantero}, \citenamefont {Mendoza}, \citenamefont {Morgan}, \citenamefont {Murakami}, \citenamefont {Nikitina}, \citenamefont {Pandola}, \citenamefont {Paprocki}, \citenamefont {Perl}, \citenamefont {Petrović}, \citenamefont {Pia}, \citenamefont {Pokorski}, \citenamefont {Quesada}, \citenamefont {Raine}, \citenamefont {Reis}, \citenamefont {Ribon}, \citenamefont {{Ristić Fira}}, \citenamefont {Romano}, \citenamefont {Russo}, \citenamefont {Santin}, \citenamefont {Sasaki}, \citenamefont {Sawkey}, \citenamefont {Shin}, \citenamefont {Strakovsky}, \citenamefont {Taborda}, \citenamefont {Tanaka}, \citenamefont {Tomé}, \citenamefont {Toshito}, \citenamefont {Tran}, \citenamefont {Truscott}, \citenamefont {Urban}, \citenamefont {Uzhinsky}, \citenamefont {Verbeke}, \citenamefont {Verderi}, \citenamefont {Wendt}, \citenamefont {Wenzel}, \citenamefont {Wright}, \citenamefont {Wright}, \citenamefont {Yamashita}, \citenamefont {Yarba},\ and\ \citenamefont {Yoshida}}]{Allison_2016}%
  \BibitemOpen
  \bibfield  {author} {\bibinfo {author} {\bibfnamefont {J.}~\bibnamefont {Allison}}, \bibinfo {author} {\bibfnamefont {K.}~\bibnamefont {Amako}}, \bibinfo {author} {\bibfnamefont {J.}~\bibnamefont {Apostolakis}}, \bibinfo {author} {\bibfnamefont {P.}~\bibnamefont {Arce}}, \bibinfo {author} {\bibfnamefont {M.}~\bibnamefont {Asai}}, \bibinfo {author} {\bibfnamefont {T.}~\bibnamefont {Aso}}, \bibinfo {author} {\bibfnamefont {E.}~\bibnamefont {Bagli}}, \bibinfo {author} {\bibfnamefont {A.}~\bibnamefont {Bagulya}}, \bibinfo {author} {\bibfnamefont {S.}~\bibnamefont {Banerjee}}, \bibinfo {author} {\bibfnamefont {G.}~\bibnamefont {Barrand}}, \bibinfo {author} {\bibfnamefont {B.}~\bibnamefont {Beck}}, \bibinfo {author} {\bibfnamefont {A.}~\bibnamefont {Bogdanov}}, \bibinfo {author} {\bibfnamefont {D.}~\bibnamefont {Brandt}}, \bibinfo {author} {\bibfnamefont {J.}~\bibnamefont {Brown}}, \bibinfo {author} {\bibfnamefont {H.}~\bibnamefont {Burkhardt}}, \bibinfo {author} {\bibfnamefont {P.}~\bibnamefont {Canal}}, \bibinfo {author} {\bibfnamefont {D.}~\bibnamefont {Cano-Ott}}, \bibinfo {author} {\bibfnamefont {S.}~\bibnamefont {Chauvie}}, \bibinfo {author} {\bibfnamefont {K.}~\bibnamefont {Cho}}, \bibinfo {author} {\bibfnamefont {G.}~\bibnamefont {Cirrone}}, \bibinfo {author} {\bibfnamefont {G.}~\bibnamefont {Cooperman}}, \bibinfo {author} {\bibfnamefont {M.}~\bibnamefont {Cortés-Giraldo}}, \bibinfo {author} {\bibfnamefont {G.}~\bibnamefont {Cosmo}}, \bibinfo {author} {\bibfnamefont {G.}~\bibnamefont {Cuttone}}, \bibinfo {author} {\bibfnamefont {G.}~\bibnamefont {Depaola}}, \bibinfo {author} {\bibfnamefont {L.}~\bibnamefont {Desorgher}}, \bibinfo {author} {\bibfnamefont {X.}~\bibnamefont {Dong}}, \bibinfo {author} {\bibfnamefont {A.}~\bibnamefont {Dotti}}, \bibinfo {author} {\bibfnamefont {V.}~\bibnamefont {Elvira}}, \bibinfo {author} {\bibfnamefont {G.}~\bibnamefont {Folger}}, \bibinfo {author} {\bibfnamefont {Z.}~\bibnamefont {Francis}}, \bibinfo {author} {\bibfnamefont {A.}~\bibnamefont {Galoyan}}, \bibinfo {author} {\bibfnamefont {L.}~\bibnamefont {Garnier}}, \bibinfo {author} {\bibfnamefont {M.}~\bibnamefont {Gayer}}, \bibinfo {author} {\bibfnamefont {K.}~\bibnamefont {Genser}}, \bibinfo {author} {\bibfnamefont {V.}~\bibnamefont {Grichine}}, \bibinfo {author} {\bibfnamefont {S.}~\bibnamefont {Guatelli}}, \bibinfo {author} {\bibfnamefont {P.}~\bibnamefont {Guèye}}, \bibinfo {author} {\bibfnamefont {P.}~\bibnamefont {Gumplinger}}, \bibinfo {author} {\bibfnamefont {A.}~\bibnamefont {Howard}}, \bibinfo {author} {\bibfnamefont {I.}~\bibnamefont {Hřivnáčová}}, \bibinfo {author} {\bibfnamefont {S.}~\bibnamefont {Hwang}}, \bibinfo {author} {\bibfnamefont {S.}~\bibnamefont {Incerti}}, \bibinfo {author} {\bibfnamefont {A.}~\bibnamefont {Ivanchenko}}, \bibinfo {author} {\bibfnamefont {V.}~\bibnamefont {Ivanchenko}}, \bibinfo {author} {\bibfnamefont {F.}~\bibnamefont {Jones}}, \bibinfo {author} {\bibfnamefont {S.}~\bibnamefont {Jun}}, \bibinfo {author} {\bibfnamefont {P.}~\bibnamefont {Kaitaniemi}}, \bibinfo {author} {\bibfnamefont {N.}~\bibnamefont {Karakatsanis}}, \bibinfo {author} {\bibfnamefont {M.}~\bibnamefont {Karamitros}}, \bibinfo {author} {\bibfnamefont {M.}~\bibnamefont {Kelsey}}, \bibinfo {author} {\bibfnamefont {A.}~\bibnamefont {Kimura}}, \bibinfo {author} {\bibfnamefont {T.}~\bibnamefont {Koi}}, \bibinfo {author} {\bibfnamefont {H.}~\bibnamefont {Kurashige}}, \bibinfo {author} {\bibfnamefont {A.}~\bibnamefont {Lechner}}, \bibinfo {author} {\bibfnamefont {S.}~\bibnamefont {Lee}}, \bibinfo {author} {\bibfnamefont {F.}~\bibnamefont {Longo}}, \bibinfo {author} {\bibfnamefont {M.}~\bibnamefont {Maire}}, \bibinfo {author} {\bibfnamefont {D.}~\bibnamefont {Mancusi}}, \bibinfo {author} {\bibfnamefont {A.}~\bibnamefont {Mantero}}, \bibinfo {author} {\bibfnamefont {E.}~\bibnamefont {Mendoza}}, \bibinfo {author} {\bibfnamefont {B.}~\bibnamefont {Morgan}}, \bibinfo {author} {\bibfnamefont {K.}~\bibnamefont {Murakami}}, \bibinfo {author} {\bibfnamefont {T.}~\bibnamefont {Nikitina}}, \bibinfo {author} {\bibfnamefont {L.}~\bibnamefont {Pandola}}, \bibinfo {author} {\bibfnamefont {P.}~\bibnamefont {Paprocki}}, \bibinfo {author} {\bibfnamefont {J.}~\bibnamefont {Perl}}, \bibinfo {author} {\bibfnamefont {I.}~\bibnamefont {Petrović}}, \bibinfo {author} {\bibfnamefont {M.}~\bibnamefont {Pia}}, \bibinfo {author} {\bibfnamefont {W.}~\bibnamefont {Pokorski}}, \bibinfo {author} {\bibfnamefont {J.}~\bibnamefont {Quesada}}, \bibinfo {author} {\bibfnamefont {M.}~\bibnamefont {Raine}}, \bibinfo {author} {\bibfnamefont {M.}~\bibnamefont {Reis}}, \bibinfo {author} {\bibfnamefont {A.}~\bibnamefont {Ribon}}, \bibinfo {author} {\bibfnamefont {A.}~\bibnamefont {{Ristić Fira}}}, \bibinfo {author} {\bibfnamefont {F.}~\bibnamefont {Romano}}, \bibinfo {author} {\bibfnamefont {G.}~\bibnamefont {Russo}}, \bibinfo {author} {\bibfnamefont {G.}~\bibnamefont {Santin}}, \bibinfo {author} {\bibfnamefont {T.}~\bibnamefont {Sasaki}}, \bibinfo {author} {\bibfnamefont {D.}~\bibnamefont {Sawkey}}, \bibinfo {author} {\bibfnamefont {J.}~\bibnamefont {Shin}}, \bibinfo {author} {\bibfnamefont {I.}~\bibnamefont {Strakovsky}}, \bibinfo {author} {\bibfnamefont {A.}~\bibnamefont {Taborda}}, \bibinfo {author} {\bibfnamefont {S.}~\bibnamefont {Tanaka}}, \bibinfo {author} {\bibfnamefont {B.}~\bibnamefont {Tomé}}, \bibinfo {author} {\bibfnamefont {T.}~\bibnamefont {Toshito}}, \bibinfo {author} {\bibfnamefont {H.}~\bibnamefont {Tran}}, \bibinfo {author} {\bibfnamefont {P.}~\bibnamefont {Truscott}}, \bibinfo {author} {\bibfnamefont {L.}~\bibnamefont {Urban}}, \bibinfo {author} {\bibfnamefont {V.}~\bibnamefont {Uzhinsky}}, \bibinfo {author} {\bibfnamefont {J.}~\bibnamefont {Verbeke}}, \bibinfo {author} {\bibfnamefont {M.}~\bibnamefont {Verderi}}, \bibinfo {author} {\bibfnamefont {B.}~\bibnamefont {Wendt}}, \bibinfo {author} {\bibfnamefont {H.}~\bibnamefont {Wenzel}}, \bibinfo {author} {\bibfnamefont {D.}~\bibnamefont {Wright}}, \bibinfo {author} {\bibfnamefont {D.}~\bibnamefont {Wright}}, \bibinfo {author} {\bibfnamefont {T.}~\bibnamefont {Yamashita}}, \bibinfo {author} {\bibfnamefont {J.}~\bibnamefont {Yarba}},\ and\ \bibinfo {author} {\bibfnamefont {H.}~\bibnamefont {Yoshida}},\ }\bibfield  {title} {\bibinfo {title} {{Recent} developments in {Geant4}},\ }\href {https://doi.org/https://doi.org/10.1016/j.nima.2016.06.125} {\bibfield  {journal} {\bibinfo  {journal} {Nucl. Instrum. Meth. A.}\ }\textbf {\bibinfo {volume} {835}},\ \bibinfo {pages} {186} (\bibinfo {year} {2016})}\BibitemShut {NoStop}%
\bibitem [{\citenamefont {Weller}\ \emph {et~al.}(2015)\citenamefont {Weller}, \citenamefont {Ahmed},\ and\ \citenamefont {Wu}}]{Weller2015}%
  \BibitemOpen
  \bibfield  {author} {\bibinfo {author} {\bibfnamefont {H.~R.}\ \bibnamefont {Weller}}, \bibinfo {author} {\bibfnamefont {M.~W.}\ \bibnamefont {Ahmed}},\ and\ \bibinfo {author} {\bibfnamefont {Y.~K.}\ \bibnamefont {Wu}},\ }\bibfield  {title} {\bibinfo {title} {{Nuclear Physics Research at the High Intensity Gamma-Ray Source (HI$\gamma$S)}},\ }\href {https://doi.org/10.1080/10619127.2015.1035932} {\bibfield  {journal} {\bibinfo  {journal} {Nucl. Phys. News}\ }\textbf {\bibinfo {volume} {25}},\ \bibinfo {pages} {19} (\bibinfo {year} {2015})}\BibitemShut {NoStop}%
\bibitem [{\citenamefont {Hidding}\ \emph {et~al.}(2019)\citenamefont {Hidding}, \citenamefont {Beaton}, \citenamefont {Boulton}, \citenamefont {Corde}, \citenamefont {Doepp}, \citenamefont {Habib}, \citenamefont {Heinemann}, \citenamefont {Irman}, \citenamefont {Karsch}, \citenamefont {Kirwan}, \citenamefont {Knetsch}, \citenamefont {Manahan}, \citenamefont {de~la Ossa}, \citenamefont {Nutter}, \citenamefont {Scherkl}, \citenamefont {Schramm},\ and\ \citenamefont {Ullmann}}]{Hidding2019}%
  \BibitemOpen
  \bibfield  {author} {\bibinfo {author} {\bibfnamefont {B.}~\bibnamefont {Hidding}}, \bibinfo {author} {\bibfnamefont {A.}~\bibnamefont {Beaton}}, \bibinfo {author} {\bibfnamefont {L.}~\bibnamefont {Boulton}}, \bibinfo {author} {\bibfnamefont {S.}~\bibnamefont {Corde}}, \bibinfo {author} {\bibfnamefont {A.}~\bibnamefont {Doepp}}, \bibinfo {author} {\bibfnamefont {F.~A.}\ \bibnamefont {Habib}}, \bibinfo {author} {\bibfnamefont {T.}~\bibnamefont {Heinemann}}, \bibinfo {author} {\bibfnamefont {A.}~\bibnamefont {Irman}}, \bibinfo {author} {\bibfnamefont {S.}~\bibnamefont {Karsch}}, \bibinfo {author} {\bibfnamefont {G.}~\bibnamefont {Kirwan}}, \bibinfo {author} {\bibfnamefont {A.}~\bibnamefont {Knetsch}}, \bibinfo {author} {\bibfnamefont {G.~G.}\ \bibnamefont {Manahan}}, \bibinfo {author} {\bibfnamefont {A.~M.}\ \bibnamefont {de~la Ossa}}, \bibinfo {author} {\bibfnamefont {A.}~\bibnamefont {Nutter}}, \bibinfo {author} {\bibfnamefont {P.}~\bibnamefont {Scherkl}}, \bibinfo {author} {\bibfnamefont {U.}~\bibnamefont {Schramm}},\ and\ \bibinfo {author} {\bibfnamefont {D.}~\bibnamefont {Ullmann}},\ }\bibfield  {title} {\bibinfo {title} {{F}undamentals and {A}pplications of {H}ybrid {LWFA}-{PWFA}},\ }in\ \href {https://doi.org/10.1007/978-3-030-25850-4_5} {\emph {\bibinfo {booktitle} {Springer Proceedings in Physics}}}\ (\bibinfo  {publisher} {Springer International Publishing},\ \bibinfo {year} {2019})\ pp.\ \bibinfo {pages} {95--120}\BibitemShut {NoStop}%
\bibitem [{awa()}]{awake}%
  \BibitemOpen
  \href {home.cern/science/accelerators/awake} {\bibinfo {title} {The {A}dvanced {P}roton {D}riven {P}lasma {W}akefield {A}cceleration {E}xperiment (awake)}},\ \bibinfo {howpublished} {https://home.cern/science/accelerators/awake}\BibitemShut {NoStop}%
\bibitem [{\citenamefont {Shen}\ and\ \citenamefont {Zhang}(2011)}]{Shen2011}%
  \BibitemOpen
  \bibfield  {author} {\bibinfo {author} {\bibfnamefont {B.}~\bibnamefont {Shen}}\ and\ \bibinfo {author} {\bibfnamefont {X.}~\bibnamefont {Zhang}},\ }\bibfield  {title} {\bibinfo {title} {{Generation of energetic protons from GeV to TeV}},\ }in\ \href {https://doi.org/10.1117/12.889154} {\emph {\bibinfo {booktitle} {Laser Acceleration of Electrons, Protons, and Ions; and Medical Applications of Laser-Generated Secondary Sources of Radiation and Particles}}},\ Vol.\ \bibinfo {volume} {8079},\ \bibinfo {editor} {edited by\ \bibinfo {editor} {\bibfnamefont {W.~P.}\ \bibnamefont {Leemans}}, \bibinfo {editor} {\bibfnamefont {E.}~\bibnamefont {Esarey}}, \bibinfo {editor} {\bibfnamefont {S.~M.}\ \bibnamefont {Hooker}},\ and\ \bibinfo {editor} {\bibfnamefont {K.~W.~D.}\ \bibnamefont {Ledingham}}},\ \bibinfo {organization} {International Society for Optics and Photonics}\ (\bibinfo  {publisher} {SPIE},\ \bibinfo {year} {2011})\ pp.\ \bibinfo {pages} {128 -- 139}\BibitemShut {NoStop}%
\bibitem [{\citenamefont {Zheng}\ \emph {et~al.}(2012)\citenamefont {Zheng}, \citenamefont {Wang}, \citenamefont {Yan}, \citenamefont {Tajima}, \citenamefont {Yu},\ and\ \citenamefont {He}}]{Zheng2012}%
  \BibitemOpen
  \bibfield  {author} {\bibinfo {author} {\bibfnamefont {F.~L.}\ \bibnamefont {Zheng}}, \bibinfo {author} {\bibfnamefont {H.~Y.}\ \bibnamefont {Wang}}, \bibinfo {author} {\bibfnamefont {X.~Q.}\ \bibnamefont {Yan}}, \bibinfo {author} {\bibfnamefont {T.}~\bibnamefont {Tajima}}, \bibinfo {author} {\bibfnamefont {M.~Y.}\ \bibnamefont {Yu}},\ and\ \bibinfo {author} {\bibfnamefont {X.~T.}\ \bibnamefont {He}},\ }\bibfield  {title} {\bibinfo {title} {Sub-{TeV} proton beam generation by ultra-intense laser irradiation of foil-and-gas target},\ }\href {https://doi.org/10.1063/1.3684658} {\bibfield  {journal} {\bibinfo  {journal} {Phys. Plasmas}\ }\textbf {\bibinfo {volume} {19}},\ \bibinfo {pages} {023111} (\bibinfo {year} {2012})}\BibitemShut {NoStop}%
\end{thebibliography}%

\end{document}